\PassOptionsToPackage{dvipsnames}{xcolor}
\documentclass[acmsmall,screen,nonacm]{acmart}
\usepackage{amsmath,amssymb,mathtools}
\usepackage{booktabs,array,tabularx,longtable}
\usepackage{enumitem}
\usepackage{graphicx}
\usepackage{subcaption}
\usepackage{listings}
\lstdefinestyle{paperprogram}{
  basicstyle=\small\ttfamily,
  columns=fullflexible,
  keepspaces=true,
  showstringspaces=false,
  aboveskip=5pt,
  belowskip=5pt,
  frame=tb,
  framerule=0.4pt,
  framesep=5pt,
  escapeinside={(*@}{@*)}
}
\usepackage{algorithm,algpseudocode}
\usepackage{booktabs,graphicx,tikz}
\usepackage{siunitx}
\usepackage{thm-restate}

\makeatletter
\@ifundefined{newcounteralias}{}{%
  \renewcommand\thmt@autorefsetup{%
    \@xa\def\csname\thmt@envname autorefname\@xa\endcsname
      \@xa{\thmt@thmname}%
  }%
}
\makeatother

\usepackage[nameinlink,noabbrev]{cleveref}

\usepackage{tikz}
\usetikzlibrary{arrows.meta,calc,positioning,fit}

\newtheorem{theorem}{Theorem}

\theoremstyle{acmdefinition}

\theoremstyle{acmplain}
\newtheorem{problem}[theorem]{Problem}

\crefname{lemma}{lemma}{lemmas}
\crefname{proposition}{proposition}{propositions}
\crefname{corollary}{corollary}{corollaries}
\crefname{problem}{problem}{problems}

\newcommand{\name}{\textsc{CoEst}}
\newcommand{\pairedestimation}{\textsc{PairedEstimation}}
\newcommand{\iscanonical}{\textsc{IsCanonical}}
\newcommand{\mazsampler}{\textsc{T}}

\newcommand{\pheading}[1]{%
  \par\medskip
  \noindent\textbf{#1}\enspace\ignorespaces
}
\newcommand{\nats}{\mathbb{N}}
\newcommand{\RF}{\mathrm{RF}}
\newcommand{\Maz}{\mathrm{Maz}}
\newcommand{\CRF}{C_{\RF}}

\newcommand{\sharpP}{\mathsf{\#P}}

\newcommand{\po}{\mathit{po}}
\newcommand{\rf}{\mathit{rf}}
\newcommand{\co}{\mathit{co}}
\newcommand{\fr}{\mathit{fr}}

\newcommand{\Exec}{\operatorname{Exec}}

\newcommand{\Realizable}{\operatorname{Realizable}_{\mathrm{SC}}}

\DeclareMathOperator{\acyclic}{acyclic}

\title{Quantifying the Advantage of Coarser DPOR}

\author{Minjian Zhang}
\affiliation{%
  \institution{Nanyang Technological University}
  \city{Singapore}
  \country{Singapore}}
\email{minjian.zhang@ntu.edu.sg}

\author{Rupak Majumdar}
\affiliation{%
  \institution{Max Planck Institute for Software Systems}
  \city{Kaiserslautern}
  \country{Germany}}
\email{rupak@mpi-sws.org}

\author{Umang Mathur}
\affiliation{%
  \institution{National University of Singapore}
  \city{Singapore}
  \country{Singapore}}
\email{umathur@nus.edu.sg}

\author{C.-H. Luke Ong}
\affiliation{%
  \institution{Nanyang Technological University}
  \city{Singapore}
  \country{Singapore}}
\email{luke.ong@ntu.edu.sg}

\renewcommand{\shortauthors}{Zhang, Majumdar, Mathur, and Ong}
\authorsaddresses{%
  Authors' contact information:
  Minjian Zhang, \href{mailto:minjian.zhang@ntu.edu.sg}{\nolinkurl{minjian.zhang@ntu.edu.sg}};
  Rupak Majumdar, \href{mailto:rupak@mpi-sws.org}{\nolinkurl{rupak@mpi-sws.org}};
  Umang Mathur, \href{mailto:umathur@nus.edu.sg}{\nolinkurl{umathur@nus.edu.sg}};
  C.-H. Luke Ong, \href{mailto:luke.ong@ntu.edu.sg}{\nolinkurl{luke.ong@ntu.edu.sg}}}

\begin{document}
\setlength{\textfloatsep}{10pt plus 2pt minus 2pt}
\raggedbottom
\begin{abstract}
Dynamic partial-order reduction (DPOR) has emerged as a promising approach
to addressing interleaving explosion in bounded model checking of
concurrent programs. Given an equivalence relation on interleavings,
the model checker explores a program's executions while minimizing
repeated visits to the same class; \emph{optimal} model checkers avoid
such repetition altogether. A coarser equivalence promises faster
exploration by reducing the number of classes. This has motivated
alternatives to Mazurkiewicz's trace equivalence such as reads-from
and reads-value-from equivalence. 
This promise, however, does not always materialize in practice: while a coarser equivalence can yield exponentially fewer classes on some families of programs, it may offer little or no reduction for the program at hand.

In this work, we show how to quantify the advantage of coarser
equivalences for individual programs. We focus on the problem of counting the reads-from
(RF) equivalence classes induced by a given program, as well as the problem of  computing the ratio of RF to Mazurkiewicz class counts
under sequential consistency. We show that both problems are
$\sharpP$-hard and that the reduction ratio remains hard to approximate
accurately under standard complexity assumptions.

Guided by these limits, we present \name{}, a sampling-based algorithm
for estimating both the RF count and its ratio to the Mazurkiewicz
count. For a fixed number of threads, each trial runs in polynomial
time and yields unbiased count estimates. As the number of independent
trials grows, the resulting count and ratio estimates converge almost
surely to their true values. Our algorithm can also provide statistical
evidence for deciding whether Mazurkiewicz or RF model checking is
likely to be faster for a given program. In experiments across 34
benchmark configurations, our method selects the faster checker in
99.4\% of runs using a one-minute estimation budget.
\end{abstract}

\maketitle

\section{Introduction}
\label{sec:introduction}

When writing concurrent programs, developers must reason about all
possible interleavings of operations from different threads. This makes
concurrent software especially difficult to get right. Subtle concurrency
errors routinely find their way into deployed software such as
libraries that implement concurrent data structures and
synchronization mechanisms. Worse still, these errors are often difficult
to reproduce and debug. Automated tools like (bounded) model checkers
offer a first line of defense against concurrency errors by
systematically exploring program
executions, say upto a bounded length, to find violations of correctness properties.

The number of possible interleavings of a program can unfortunately grow
exponentially with program size, making exhaustive exploration
prohibitively expensive. In response, modern model checkers employ
dynamic partial-order reduction (DPOR)~\cite{flanagan2005dpor}.
In essence, DPOR-based model
checkers group executions into equivalence classes and explore
representative executions from each class. This reduces the state space
by avoiding redundant exploration. Model checkers such as
Nidhugg~\cite{rfsc2019} and GenMC~\cite{genmc2021}
(using TruSt~\cite{TruSt2022}) additionally support \emph{optimal} exploration,
visiting exactly one representative per equivalence class.

The choice of equivalence relation determines how much DPOR can reduce
exploration. Classical DPOR uses Mazurkiewicz equivalence, which groups
executions that differ only in the order of independent operations.
To reduce exploration further, researchers have developed coarser
equivalences that merge more executions into each class. Reads-from (RF)
equivalence, for example, groups executions with the same events and
read sources, even when they order conflicting writes differently.
Under sequential consistency, this can leave exponentially fewer classes
to explore~\cite{dcpor2018,rfsc2019}. More broadly, researchers have
proposed coarser equivalences that can yield exponentially fewer classes
than earlier approaches~\cite{observers2018,vcpor2019,rvf2021,farzan2024coarser}.
But does this theoretical advantage translate into faster model checking?
This is the central question we seek to answer in this paper,
while focusing on Mazurkiewicz and RF equivalence, under the
sequential consistency memory model.

\begin{figure}[tbp]
  \centering
  \captionsetup[subfigure]{position=top,font=small,
    labelfont=bf,skip=6pt}
  \begin{subfigure}[t]{0.47\linewidth}
    \caption{$P_1$: unobserved writes}
    \label{fig:unobserved-writes}
\begin{lstlisting}[style=paperprogram]
Initially: x = 0

Thread i in {1, ..., n}:
    x := 1;
\end{lstlisting}
  \end{subfigure}\hfill
  \begin{subfigure}[t]{0.47\linewidth}
    \caption{$P_2$: locked list insertion}
    \label{fig:locked-list-insertion}
\begin{lstlisting}[style=paperprogram]
Initially: head = null

Thread i in {1, ..., n}:
    lock(m);
    old := head;
    node[i].next := old;
    head := &node[i];
    unlock(m);
\end{lstlisting}
  \end{subfigure}
  \caption{Each thread executes the shown code once.
    \subref{fig:unobserved-writes} Unobserved writes:
    $n!$ Mazurkiewicz classes, one RF class.
    \subref{fig:locked-list-insertion} List insertion:
    $n!$ classes under both equivalences.
    Nodes are distinct and preallocated; the mutex is abstract and
    blocking. Counts include only the displayed data accesses.}
  \label{fig:counting-extremes}
  \Description{Programs P1 and P2 run by n threads.
    P1 has a single store per thread, producing n factorial
    Mazurkiewicz classes and one RF class.
    P2 has a serialized linked-list insertion per thread,
    producing n factorial classes under both equivalences.}
\end{figure}

The answer is not straightforward. First, the exponential advantage may
not show up in all programs. Consider, for example, the two programs
$P_1$ and $P_2$ in Figure~\ref{fig:counting-extremes}.
$P_1$ has one RF class but $n!$ Mazurkiewicz classes.
$P_2$, on the other hand, has $n!$ classes under both equivalences.
Second, even for programs for which coarsening offers an advantage in
terms of the number of classes, a model checker may
still perform more work when generating representatives
of the coarser classes and, in turn, may spend overall more
time despite actually exploring fewer classes.
In this paper, we take a quantitative view of this tradeoff
and ask if we can estimate the advantage of a
coarser equivalence for a given program before completing its exploration.
Such estimates would help us choose which model checker to run.
For a given program, suppose exploring an RF
class costs, on average, $h$ times as much as exploring a Mazurkiewicz
class. Let $\rho$ denote the ratio of the number of RF classes to the
number of Mazurkiewicz classes. The total exploration time for RF is
then approximately $h\rho$ times that for Mazurkiewicz, so RF is expected
to be faster when $\rho<1/h$. We therefore ask: \emph{can we estimate $\rho$
without completing the exploration?}

\begin{samepage}
\pheading{The counting complexity landscape.}
We first study the problem from a complexity-theoretic perspective.
We begin with some background on the complexity of counting equivalence
classes. Counting Mazurkiewicz classes was recently shown to be complete
for the class $\sharpP$~\cite{pldi26}.\footnote{$\sharpP$ is the collection
of those counting problems $f : \Sigma^* \to \nats$ for which there is a non-deterministic
polynomial time Turing Machine $M_f$ such that for each input $x$,
the number of accepting paths of $M_f$ is exactly $f(x)$.
A canonical $\sharpP$-complete problem is $\#\mathrm{SAT}$, which asks to compute
the number of satisfying assignments for a given propositional formula presented in 3-CNF. 
An efficient exact algorithm for any $\sharpP$-hard problem would also yield one for $\#\mathrm{SAT}$;
thus, $\sharpP$-hardness provides evidence of computational intractability.}
In this work, we study two concrete problems:
\par
\end{samepage}

\begin{restatable}[Counting RF classes]{problem}{rfcountingproblem}
\label{prob:rf-counting}
\label{prob:counting-execution-classes}
Given a bounded SC program $P$, compute $C_{\RF}(P)$, the number of
RF-equivalence classes of its maximal executions.
\end{restatable}

\begin{restatable}[Computing the reduction ratio]{problem}{reductionratioproblem}
\label{prob:reduction-ratio}
Given such a program $P$, compute
\[
  \rho(P)=\frac{C_{\RF}(P)}{C_{\Maz}(P)},
\]
where $C_{\Maz}(P)$ counts its Mazurkiewicz classes.
\end{restatable}

\begingroup
\clubpenalties 3 10000 10000 0
\widowpenalties 3 10000 10000 0
\displaywidowpenalty=10000
We show that both problems above are harder than the problem of counting Mazurkiewicz classes, and thus also $\sharpP$-hard. When the number of threads
is part of the input, we further show that RF counting does not belong
to $\sharpP$ unless $\mathrm{NP}\subseteq\mathrm{SPP}$.\footnote{\label{fn:complexity-assumptions}At
the time of writing, none of the complexity-theoretic assumptions used
in this paper has been proved or disproved.} Under the
assumption that this inclusion does not hold, coarsening can therefore
move counting outside $\sharpP$, despite leaving fewer classes.

Given this evidence of intractability of \emph{exact} computation,
we next consider a natural question: can we efficiently
\emph{approximate} $C_{\RF}(P)$ or $\rho(P)$ to a prescribed relative
error, with high probability?
We first remark that, even for Mazurkiewicz counting,
such a guarantee is impossible in polynomial time (unless
$\mathrm{NP}=\mathrm{RP}$)\footref{fn:complexity-assumptions}~\cite{pldi26}. Motivated by this
limitation, in~\cite{pldi26}, Balasubramanian et al. pursued a weaker goal:
constructing a polynomial-time \emph{unbiased estimator} for
$C_{\Maz}(P)$.\footnote{An unbiased estimator for a counting
function $f$ is a randomized algorithm $A$ whose expected output
on every input $x$ equals $f(x)$. Unbiasedness alone imposes no
bound on the variance of its output. We call the estimator
polynomial-time if each trial runs in polynomial time in the
input size.}
Their result raises the question: can we obtain the same guarantee
for RF counting, or estimate the reduction ratio directly?
Unfortunately, we show that neither quantity admits a
polynomial-time unbiased estimator when the number of threads
is part of the input, assuming
$\mathrm{PP}^{\mathrm{NP}}\neq\mathrm{PP}$.\footref{fn:complexity-assumptions}
We also show that $\rho(P)$ admits no fully polynomial randomized
approximation scheme (FPRAS) unless $\mathrm{NP}=\mathrm{RP}$,
even for loop-free programs with two threads and Boolean shared
variables.\footref{fn:complexity-assumptions}
On a positive note, we show that when we fix the number of threads (i.e., treat it as a constant), 
then RF counting belongs to $\sharpP$ and therefore admits an unbiased estimator with
polynomial time per trial.

\pheading{A practical estimation algorithm.}
Despite these worst-case barriers, we propose a practical algorithm,
\name{} (for ``Coarsening Estimator''), for estimating $C_{\RF}(P)$
and $\rho(P)$. \name{} is a randomized algorithm, and each trial
produces unbiased estimates of $C_{\RF}(P)$ and $C_{\Maz}(P)$.
It guarantees that (1) each trial runs in polynomial time when the
number of threads is fixed, and (2) it is a strongly consistent
estimator of $\rho(P)$, i.e., the ratio of the average RF and
Mazurkiewicz count estimates converges almost surely to $\rho(P)$
as the number of independent trials grows. For any finite number
$N$ of trials, we also bound the probability that the relative error
exceeds a given tolerance, in terms of the true counts and variances.

At a high level, borrowing insights from prior work on
Mazurkiewicz counting~\cite{pldi26} central to making estimation practical,
\name{} also works atop the optimal model checker TruST (with Mazurkiewicz equivalence based DPOR)
and uses Knuth's estimator~\cite{knuth1975} to sample leaves from its
exploration tree. To adapt this approach to RF counting, we need to
avoid overcounting; several Mazurkiewicz classes may belong to the
same RF class. We do this by fixing, a priori, exactly one canonical
Mazurkiewicz representative per RF class. We develop a procedure,
\iscanonical{}, that checks whether a given execution graph is this
representative. When sampling, we assign a leaf weight of one if it
represents a canonical maximal execution, and zero otherwise. Using
these modified weights in Knuth's estimator gives an unbiased
RF-count estimate. Alongside, we also maintain the original Mazurkiewicz
contributions and use the two accumulated estimates to compute the
reduction ratio.

Finally, we show how we can use \name{}'s samples to help choose between RF and
Mazurkiewicz model checkers. Recall that RF exploration is expected
to be faster when $\rho(P)<1/h$. We use a statistical test to assess
whether the samples provide sufficient evidence that $\rho(P)$ lies
below or above this threshold. The test uses a normal approximation,
with an asymptotic significance guarantee. When the evidence is
insufficient, it reports an inconclusive result.

\pheading{Empirical Evaluation.}
\looseness=-1
We implement \name{} in GenMC and evaluate whether short
estimation runs help select the faster checker and whether
the savings outweigh the cost of making this choice.
Our benchmarks include concurrent data structures,
synchronization algorithms, a simplified device-driver model,
and synthetic programs.

We use off-the-shelf model checkers implementing
Mazurkiewicz-based and RF-based DPOR. We run each checker for up to one minute to measure its
exploration rate, obtaining an estimate of the relative
per-class cost $h$ that we combine with \name{}'s estimate
of $\rho(P)$.
With a one-minute estimation budget, population budget $b=50$,
and ten seeds per configuration, we choose the faster checker
in 338 of 340 decisions (99.4\%) across 34 configurations. On the 25 configurations where neither pilot finishes,
the reconstructed total cost, including both pilots
and estimation, is 43.9\% lower than ideal round robin
and only 12.2\% above an oracle that selects the faster
checker at no selection cost.

We also evaluate the numerical accuracy of the ratio
estimates on 29 configurations with exact reference ratios.
After one minute of sampling with $b=50$, the mean q-error
across 27 configurations is 1.08, where a q-error of one
indicates an exact estimate and two indicates a factor-of-two error.
The remaining two configurations yield zero estimates
in all or most runs because these runs sample no canonical
representative.
Although these ratios are underestimated in most runs,
the estimates still lead to the faster checker in every run.
\begin{samepage}
\pheading{Organization.}
The rest of the paper is organized as follows.
Section~\ref{sec:semantics} presents the preliminaries.
Section~\ref{sec:motivation} motivates the reduction ratio and its
use in checker selection. Section~\ref{sec:complexity} presents our
complexity results. Section~\ref{counting_algo} describes \name{},
its accuracy guarantees, and the statistical test for checker
selection. Section~\ref{sec:experiments} presents our implementation
and evaluation, and Section~\ref{sec:related} discusses related work.
Additional proofs and variance analysis appear in the appendices.
\par
\end{samepage}
\par
\endgroup

\section{Preliminaries}
\label{sec:semantics}

\subsection{Concurrent Programs}

We consider a bounded concurrent program $P$ whose threads communicate
through reads and writes to shared locations, starting from a fixed
initial state. Each thread's local computation and control flow are
deterministic once its read results are fixed. Under sequential
consistency (SC), an execution interleaves the threads' operations in
program order, and each read observes the last preceding write to
its location. We assume a finite upper bound on execution length.

Let $\mathit{Tid}$ and $\mathit{Loc}$ be the finite sets of threads
and shared locations, and let $\mathit{Val}$ be the value domain.
A shared-memory event is a tuple $e=(t,i,\ell)$, where
$t\in\mathit{Tid}$ is its thread, $i\geq 1$ is its occurrence index
within that thread, and $\ell$ is a label $\mathsf R(x,v)$ or
$\mathsf W(x,v)$, with $x\in\mathit{Loc}$ and
$v\in\mathit{Val}$, recording the operation, location, and value.
We write $\operatorname{tid}(e)$, $\operatorname{idx}(e)$,
$\operatorname{lab}(e)$, $\operatorname{op}(e)$,
$\operatorname{loc}(e)$, and $\operatorname{val}(e)$ for these
attributes. Each location $x$ has an initialization write
$\mathsf{init}_x$ that records its initial value; let
$I=\{\mathsf{init}_x:x\in\mathit{Loc}\}$.

We write $\Exec(P)$ for the set of executions of $P$, represented
by their shared-memory events. An execution is \emph{maximal} if
it cannot be extended by another program step, and
$\Exec_{\mathrm{max}}(P)$ denotes the set of maximal executions.
Boundedness and determinism imply that these sets are finite.

\subsection{Execution Graphs and Equivalences}

We describe program semantics using execution
graphs~\cite{genmc2021,TruSt2022}: the executions allowed
under SC are represented by graphs that respect the threads'
code and satisfy the SC consistency condition.

\pheading{Execution graphs.}
A \emph{full execution graph} $G=(E,\rf,\co)$ consists of:
\begin{enumerate}
  \item a finite set $E$ of labeled events containing the
        initialization writes $I$. No two thread events have
        the same thread identifier and occurrence index, and
        each thread's events are indexed consecutively in
        program order, starting from $1$.
        Let $R(G)$ and $W(G)$ denote the reads and writes in $E$,
        respectively, with $I\subseteq W(G)$, and let $W_x(G)$
        denote the writes to location $x$;
  \item a \emph{reads-from} relation
        $\rf\subseteq W(G)\times R(G)$.
        Every read $r\in R(G)$ has exactly one source
        $w\in W(G)$ with $(w,r)\in\rf$, satisfying
        $\operatorname{loc}(w)=\operatorname{loc}(r)$ and
        $\operatorname{val}(w)=\operatorname{val}(r)$;
  \item a \emph{coherence order} $\co=\bigcup_{x\in\mathit{Loc}}\co_x$,
        where each $\co_x$ is a strict total order on $W_x(G)$
        with $\mathsf{init}_x$ as its least element.
\end{enumerate}
Here, \emph{full} refers to the inclusion of both $\rf$ and $\co$;
a full graph may also represent an execution prefix.
We write $G.E$, $G.\rf$, and $G.\co$ for its components.

The \emph{program-order} relation $\po_G$ orders each thread's
events by their indices and places all initialization writes
before all thread events:
\[
  \po_G =
  I\times(E\setminus I)
  \;\cup\;
  \left\{
    (e,f)\in(E\setminus I)^2
    \;\middle|\;
    \begin{array}{l}
      \operatorname{tid}(e)=\operatorname{tid}(f),\\
      \operatorname{idx}(e)<\operatorname{idx}(f)
    \end{array}
  \right\}.
\]
The \emph{from-read} relation
\[
  \fr_G=(G.\rf)^{-1};G.\co
\]
relates each read to every write to the same location that follows
its source in coherence order, where $;$ denotes relational
composition from left to right.
A graph is compatible with $P$'s per-thread semantics if each
thread's labeled event sequence follows its code when its reads
return the recorded values. Such a graph is SC-consistent
if and only if
\begin{equation}
  \acyclic(\po_G\cup G.\rf\cup G.\co\cup\fr_G).
  \label{eq:sc}
\end{equation}

The \emph{reads-from (RF) graph} of $G$ retains its labeled
events and read sources while forgetting coherence:
\[
  \pi_{\RF}(G)=(G.E,G.\rf).
\]
Program order remains determined by the thread identifiers
and occurrence indices in the event tuples.
For an execution $\sigma\in\Exec(P)$, let $G_\sigma$ be the
full graph it induces: reads-from records each read's source,
and coherence records the order of writes to each location
in $\sigma$. The full graphs induced by maximal executions
form the set
\[
  \mathcal M(P)
  =\{G_\sigma:\sigma\in\Exec_{\mathrm{max}}(P)\}.
\]

\pheading{Equivalence and realizability.}
Two executions are Mazurkiewicz equivalent exactly when their
full graphs coincide~\cite{pldi26}, and RF-equivalent exactly
when their RF projections coincide~\cite{dcpor2018,rfsc2019}.
Thus, Mazurkiewicz equivalence refines RF equivalence:
each RF class contains one or more Mazurkiewicz classes.

A candidate full or RF graph $H$ is \emph{realizable} in $P$,
written $\Realizable(P,H)$, if some execution of $P$ induces it.
For an RF graph compatible with the threads' code, this amounts
to asking whether it admits a coherence order satisfying
\eqref{eq:sc}. This RF consistency-checking problem is
NP-complete in general~\cite{gibbons1997}.

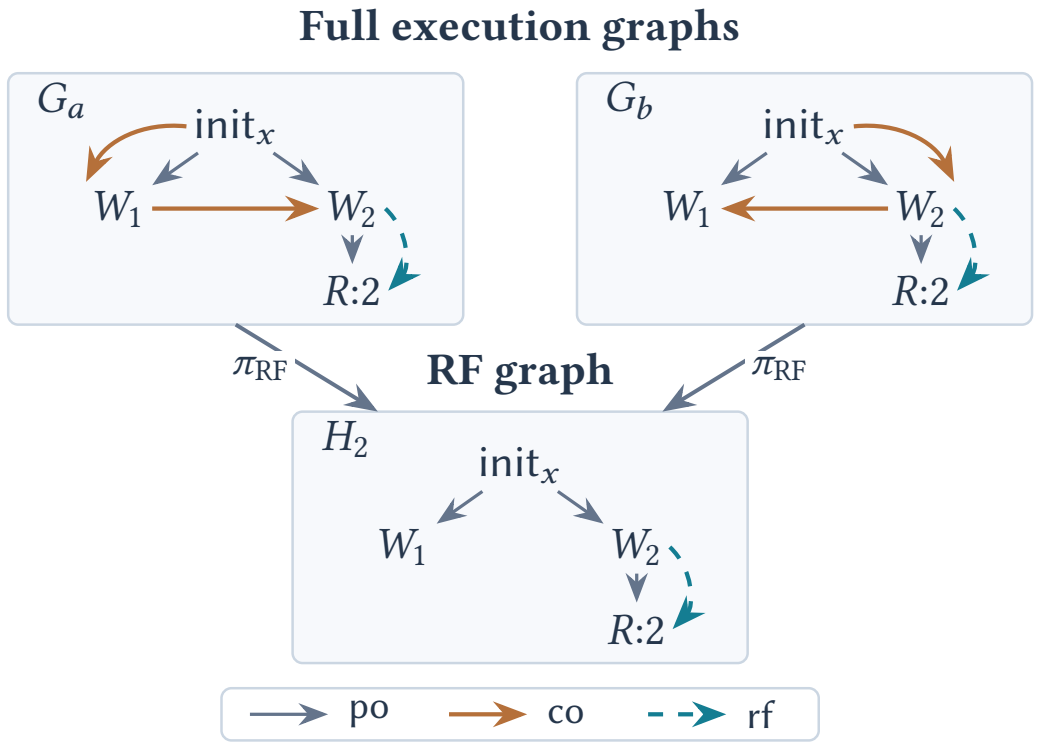
\begin{figure}[t]
\centering
\begin{minipage}[c]{0.35\linewidth}
\centering
\tiny
\textbf{Program}\par\smallskip
Initially \texttt{x = 0}\par\smallskip
\begingroup
\setlength{\tabcolsep}{5pt}
\renewcommand{\arraystretch}{1.2}
\begin{tabular}{@{}ll@{}}
\toprule
\multicolumn{1}{c}{$T_1$} & \multicolumn{1}{c}{$T_2$}\\
\midrule
$W_1:$ \texttt{x := 1;} & $W_2:$ \texttt{x := 2;}\\
                       & $R:$ \texttt{r := x;}\\
\bottomrule
\end{tabular}
\endgroup
\end{minipage}%
\hfill
\begin{minipage}[c]{0.61\linewidth}
\centering
\definecolor{prelimInk}{HTML}{25364B}
\definecolor{prelimPO}{HTML}{64748B}
\definecolor{prelimRF}{HTML}{147D92}
\definecolor{prelimCO}{HTML}{B5703A}
\definecolor{prelimBorder}{HTML}{CBD6E2}
\definecolor{prelimBG}{HTML}{F7F9FC}

\resizebox{0.8\linewidth}{!}{%
\begin{tikzpicture}[
  x=1cm,
  y=.68cm,
  font=\small,
  text=prelimInk,
  >=Stealth,
  event/.style={
    inner sep=1.2pt,
    minimum height=3.6mm
  },
  frame/.style={
    draw=prelimBorder,
    fill=prelimBG,
    rounded corners=2pt,
    line width=.5pt
  },
  poedge/.style={
    ->,draw=prelimPO,line width=.65pt
  },
  rfedge/.style={
    ->,draw=prelimRF,dashed,line width=.85pt
  },
  coedge/.style={
    ->,draw=prelimCO,line width=.85pt
  },
  projection/.style={
    ->,draw=prelimPO,line width=.8pt
  },
  edgelabel/.style={
    font=\fontsize{8}{9}\selectfont,
    inner sep=1pt,
    fill=prelimBG
  }
]

\newcommand{\smallGraph}[6]{%
  \begin{scope}[shift={(#3,#4)}]
    \draw[frame] (-1.60,-1.15) rectangle (1.60,1.45);
    \node at (-1.22,1.17) {$#2$};

    \node[event] (#1init) at (0,.90) {$\mathsf{init}_x$};
    \node[event] (#1w1) at (-.82,.05) {$W_1$};
    \node[event] (#1w2) at (.82,.05) {$W_2$};
    \node[event] (#1r) at (.82,-.80) {$R{:}#5$};

    \draw[poedge] (#1init) -- (#1w1);
    \draw[poedge] (#1init) -- (#1w2);
    \draw[poedge] (#1w2) -- (#1r);

    \draw[rfedge]
      (#1w2.east) to[bend left=45] (#1r.east);

    \ifnum#6=1
      \draw[coedge]
        (#1init.west) to[bend right=35] (#1w1.north west);
      \draw[coedge] (#1w1) -- (#1w2);
    \fi

    \ifnum#6=-1
      \draw[coedge]
        (#1init.east) to[bend left=35] (#1w2.north east);
      \draw[coedge] (#1w2) -- (#1w1);
    \fi
  \end{scope}%
}

\node[font=\small\bfseries] at (2.00,1.90)
  {Full execution graphs};
\node[font=\small\bfseries] at (2.00,-1.65)
  {RF graph};

\smallGraph{ga}{G_a}{0}{0}{2}{1}
\smallGraph{gb}{G_b}{4.00}{0}{2}{-1}
\smallGraph{h}{H_2}{2.00}{-3.50}{2}{0}

\draw[projection] (0,-1.15) --
  node[edgelabel,fill=white,left=2pt] {$\pi_{\mathrm{RF}}$}
  (1.00,-2.05);

\draw[projection] (4.00,-1.15) --
  node[edgelabel,fill=white,right=2pt] {$\pi_{\mathrm{RF}}$}
  (3.00,-2.05);

\draw[frame,fill=white]
  (-.10,-5.45) rectangle (4.10,-4.91);

\draw[poedge] (.10,-5.18) -- (.65,-5.18);
\node[edgelabel,fill=white,anchor=west]
  at (.75,-5.18) {$\mathsf{po}$};

\draw[coedge] (1.50,-5.18) -- (2.05,-5.18);
\node[edgelabel,fill=white,anchor=west]
  at (2.15,-5.18) {$\mathsf{co}$};

\draw[rfedge] (2.90,-5.18) -- (3.45,-5.18);
\node[edgelabel,fill=white,anchor=west]
  at (3.55,-5.18) {$\mathsf{rf}$};

\end{tikzpicture}%
}
\end{minipage}
\caption{A program (left) and two full execution graphs with a
common RF projection (right). $R{:}v$ records the read value;
transitive $\mathsf{po}$ and $\mathsf{co}$ edges are omitted.}
\label{fig:prelim-rf-projection}
\Description{The program on the left has two threads. Initially x
is zero. Thread one writes 1 to x; thread two writes 2 to x and
then reads x. On the right, two full graphs in the top row differ
in the coherence order of W1 and W2. Both have R reading from W2
and project down to the same RF graph. A legend below identifies
program-order, coherence, and reads-from edges.}
\end{figure}

\begin{example}[Mazurkiewicz and RF equivalence]
\label{ex:prelim-rf-program}
In Figure~\ref{fig:prelim-rf-projection}, the executions
$W_1\,W_2\,R$ and $W_2\,R\,W_1$ have different coherence orders
but the same read source. Their full graphs therefore represent
different Mazurkiewicz classes with the same RF projection,
as illustrated on the right.
The program has three Mazurkiewicz classes and two RF classes.
\end{example}

\subsection{Counting Problems}
\label{sec:counting-problems}

For $Q\in\{\Maz,\RF\}$, let $\equiv_Q$ denote the corresponding
equivalence on executions. We write
\begin{equation}
  C_Q(P)
  =\left|\Exec_{\mathrm{max}}(P)/{\equiv_Q}\right|
  \label{eq:execution-class-count}
\end{equation}
for the number of classes of maximal executions of $P$.
Thus, $C_{\Maz}(P)$ counts distinct full execution graphs, while
$C_{\RF}(P)$ counts their distinct RF projections.
With this notation, we restate the two problems introduced in
Section~\ref{sec:introduction}.

\rfcountingproblem*

\reductionratioproblem*

\subsection{Model Checking and Partial-Order Reduction}

We focus on execution-based model checkers, which explore
program behaviors by executing or interpreting the program
and check properties such as the absence of local assertion
violations. Dynamic partial-order reduction (DPOR) reduces
this exploration by grouping equivalent executions and exploring
representatives. Classical DPOR uses Mazurkiewicz equivalence
to avoid redundant interleavings of independent operations.
Under sequential consistency, RF-based DPOR uses the coarser
reads-from equivalence, so one representative can cover several
Mazurkiewicz classes~\cite{dcpor2018,rfsc2019}.

An exploration algorithm is \emph{optimal} for a chosen
equivalence if complete exploration visits exactly one
representative of each class of maximal executions.
TruSt, implemented in GenMC, provides optimal DPOR
exploration for Mazurkiewicz equivalence under
SC~\cite{TruSt2022}.
Nidhugg also supports optimal Mazurkiewicz exploration
using a different algorithm from TruSt, while its
RF-SC mode provides optimal exploration for RF
equivalence~\cite{rfsc2019}.

GenMC and Nidhugg are stateless model checking tools:
they avoid maintaining a global set of visited program states.
Statelessness does not, however, imply a polynomial bound
on exploration bookkeeping.
TruSt additionally guarantees exploration space polynomial
in the maximum execution length~\cite{TruSt2022}.

The chosen equivalence also affects the work needed to generate
each representative. For a full execution graph with a fixed
coherence order, SC consistency can be checked in polynomial
time using the acyclicity condition in~\eqref{eq:sc}.
For an RF graph, the coherence order is unspecified,
and consistency requires finding a coherence completion
satisfying that condition, an NP-complete problem in
general~\cite{gibbons1997,rfsc2019}.
A coarser equivalence can therefore reduce the number of
representatives while increasing the cost of generating them.

\section{When is Coarser DPOR Better?}
\label{sec:motivation}

RF equivalence has motivated optimal model checkers that can
explore substantially fewer executions than checkers
based on Mazurkiewicz equivalence~\cite{dcpor2018,rfsc2019}.
Since RF equivalence is coarser, one RF class can contain
many Mazurkiewicz classes.
On some programs, this reduces the number of executions
to explore by an exponential factor and leads to
substantial performance improvements.

But does exploring fewer classes always make an RF
checker faster? The answer is unfortunately no: RF checkers are
not always the fastest choice.
RF exploration repeatedly invokes consistency checks
to determine whether candidate RF graphs can be realized
under SC.
This problem is NP-complete in general when the number
of threads may grow~\cite{gibbons1997}.
Specialized procedures can nevertheless handle these
checks efficiently in practice.
For example, Abdulla et al.~\cite{rfsc2019} report that
the first two polynomial-time phases of their consistency
procedure resolved every query in their evaluation,
without invoking the expensive complete procedure.

RF exploration can nevertheless incur a higher cost
per class.
For example, on \texttt{fib-bench(6)},
Abdulla et al.~\cite{rfsc2019} report that
Nidhugg/rfsc(based on RF DPOR)  takes about 1.64 times as long as
Nidhugg/optimal(based on Mazurkiewicz DPOR), although both explore the same
number of executions. Our experiments also show substantial differences
in verification time in both directions.
On \texttt{Treiber static (L3)}, the Mazurkiewicz
checker is about 7.8 times as fast as the RF checker,
whereas on \texttt{Eratosthenes SC (N14, U16)},
the RF checker is about 16.4 times as fast
(Section~\ref{sec:experiments}).

Whether this cost is worthwhile depends on how much
RF equivalence reduces the execution space.
Figure~\ref{fig:counting-extremes} illustrates two extremes.
In Figure~\ref{fig:counting-extremes}\subref{fig:unobserved-writes}, each of $n$
threads writes to the same shared location, and no
thread reads it.
The $n!$ write orders induce distinct Mazurkiewicz
classes, but all have the same RF projection.
RF exploration therefore replaces $n!$ classes with one.
In Figure~\ref{fig:counting-extremes}\subref{fig:locked-list-insertion}, each thread
inserts a node into a mutex-protected list.
Each read of \texttt{head} identifies the preceding
insertion, so the read sources determine the entire
insertion order.
Both equivalences have $n!$ classes.
Here, RF exploration obtains no reduction to compensate
for any additional cost per class.

These examples raise the question that motivates our work:
\emph{given a program, is the reduction offered by RF
equivalence worth the cost of exploiting it?}
Existing evaluations show programs on which RF exploration
is beneficial, but we would like to answer this question
before committing to a complete exploration.
Doing so requires understanding two quantities:
the relative cost of exploring a class, and the reduction
in the number of classes to explore.

A general formal characterization of this overhead
is difficult.
Model checkers differ in their exploration algorithms
and implementations. 
For example, both GenMC~\cite{genmc2021} and
Nidhugg~\cite{rfsc2019} support optimal exploration,
but they use different exploration algorithms
and can incur different costs per explored class.
Moreover, SC consistency checking is NP-complete
in general, and the cost of a consistency query
can depend strongly on the input.
In practice, however, we can summarize the relative
exploration cost of the two model checkers on a given
program by an average factor $h$: exploring an RF
class costs approximately $h$ times as much as
exploring a Mazurkiewicz class.

This factor may be supplied by the user based on
experience, or estimated from short pilot runs.
Our experiments show that short pilots provide
reasonably accurate estimates of this relative cost
on the evaluated benchmarks
(Section~\ref{sec:experiments}).

Knowing this cost alone does not tell us which checker
will finish first.
A short run tells us how fast a checker is exploring,
but not how much it has to explore.
An RF checker may visit fewer classes during the pilot
because it spends more time on each class, yet finish
earlier because its total number of classes is much smaller.
We therefore also need to estimate the reduction
offered by RF equivalence.

The second quantity, namely the reduction ratio,
measures how much RF equivalence reduces the
Mazurkiewicz execution space.
It depends only on the program and the two equivalences,
regardless of how the checkers are implemented.
Computing this ratio is itself a natural mathematical
problem, which we study by counting execution classes
under the two equivalences. Section~\ref{sec:counting-problems}
defines these counts formally in~\eqref{eq:execution-class-count}
and states the RF-counting and reduction-ratio problems
(Problems~\ref{prob:rf-counting} and~\ref{prob:reduction-ratio}).
When $\rho(P)=1$, RF equivalence eliminates no classes;
a small ratio indicates substantial reduction.

If exploring an RF class costs $h$ times as much
as exploring a Mazurkiewicz class on average,
then RF exploration takes approximately $h\rho(P)$
times as long as Mazurkiewicz exploration.
We therefore predict RF exploration to be faster when
\[
  \rho(P)<\eta,
  \qquad
  \eta=\frac{1}{h}.
\]
For example, if an RF class costs five times as much
to explore, RF exploration must visit fewer than
one fifth as many classes to compensate.

But how can we obtain this ratio without completing
the exploration?
We first study the complexity of computing and
estimating it in Section~\ref{sec:complexity}.
These results guide the design of our estimation
algorithms in Section~\ref{counting_algo}.
We then evaluate these algorithms in
Section~\ref{sec:experiments} and show how their
ratio estimates, together with overhead estimates
from short pilot runs, help us choose between
RF and Mazurkiewicz checkers.

\section{From Mazurkiewicz to RF Counting}
\label{sec:complexity}

A direct way to compute the reduction ratio is to count
RF and Mazurkiewicz classes separately.
For Mazurkiewicz counting, Balasubramanian et al.~\cite{pldi26}
show that exact counting of bounded concurrent programs\footnote{Throughout our complexity and estimation results,
we assume that execution length, the bit length of shared
and local values, and the cost of simulating each operation,
including its local computation, are polynomially bounded
in the size of the bounded program's encoding.} is $\sharpP$-complete and that
the problem admits no fully polynomial randomized
approximation scheme (FPRAS) unless
$\mathrm{RP}=\mathrm{NP}$.
Nevertheless, they obtain an unbiased estimator by
sampling Mazurkiewicz traces: each trial runs in
polynomial time and returns a value whose expectation
equals the number of Mazurkiewicz classes.

We ask whether analogous guarantees are possible for
RF equivalence.
Although RF equivalence merges Mazurkiewicz classes
and therefore leaves fewer classes to count, does this
make the counting problem easier or harder?
Can we construct an unbiased estimator with polynomial
time per trial for the RF count, or estimate the
reduction ratio directly with the same guarantee?
If we allow a relative error and probability of failure,
are there FPRAS for the ratio?

In this section, we answer these questions from a
complexity-theoretic perspective.
When the number of threads is part of the input,
we show that, under the assumptions stated below,
RF counting lies outside $\sharpP$.
The separation extends to unbiased estimation:
unlike the Mazurkiewicz count, neither the RF count
nor the reduction ratio admits an oracle-free
unbiased estimator with polynomial time per trial.
Thus, coarsening the equivalence can make counting
computationally harder despite reducing the number
of execution classes. 

We begin with a brief review of some complexity classes.

\pheading{The classes $\mathrm{SPP}$ and $\mathrm{PP}$.}
Throughout this section, hardness and completeness for
counting problems use polynomial-time Turing reductions:
a reduction may query an oracle for the target problem
and process its answers in polynomial time.

Our lower bounds also use two decision classes.
A language $L$ belongs to $\mathrm{SPP}$ if there exist
$f,g\in\sharpP$ such that
\[
  f(x)-g(x)=
  \begin{cases}
    1, & x\in L,\\
    0, & x\notin L.
  \end{cases}
\]
The class $\mathrm{PP}$ consists of languages for which
a polynomial-time probabilistic machine accepts with
probability greater than $1/2$ exactly on inputs
in the language.

The class $\mathrm{SPP}$ is low for $\mathrm{PP}$:
giving a $\mathrm{PP}$ machine an $\mathrm{SPP}$ oracle
does not increase its computational power,
so $\mathrm{PP}^{\mathrm{SPP}}=\mathrm{PP}$~\cite{fenner1994}.
Consequently,
\[
  \mathrm{NP}\subseteq\mathrm{SPP}
  \quad\Longrightarrow\quad
  \mathrm{PP}^{\mathrm{NP}}=\mathrm{PP}.
\]
For our lower bounds on unbiased estimation, we assume
$\mathrm{PP}^{\mathrm{NP}}\neq\mathrm{PP}$.
Whether this equality holds remains open at the time when this paper is written;
see Fortnow and Grochow~\cite{fortnowGrochow2011}
for a discussion.

\pheading{Exact counting.}
We first study the complexity of counting RF classes and
computing the reduction ratio exactly.
We show that both problems are $\sharpP$-hard.

\begin{theorem}[Appendix~\ref{app:rf-sharpp-hard}]
\label{thm:rf-sharpp-hard}
For bounded deterministic SC read/write programs with
polynomially bounded execution length, value size, and
per-operation simulation cost, counting RF-equivalence
classes of maximal executions is $\sharpP$-hard under
polynomial-time Turing reductions.
The numbers of threads, shared locations, and values
may grow with the input.
\end{theorem}

\pheading{Exact computation of the reduction ratio.}
Given that RF counting is hard, we now query if we can
compute the reduction ratio efficiently without
counting the two spaces separately?
We show that exact computation of the ratio is also
$\sharpP$-hard, even for loop-free programs.

\begin{proposition}[Appendix~\ref{app:ratio-hardness}]
\label{thm:ratio-hardness}
Exact computation of
\[
  \rho(P)=\frac{C_{\RF}(P)}{C_{\Maz}(P)}
\]
is $\sharpP$-hard under polynomial-time Turing reductions,
even for loop-free deterministic SC read/write programs.
\end{proposition}

\pheading{Separation from Mazurkiewicz counting.}
The results so far show that counting RF classes and computing
the reduction ratio are at least as hard as Mazurkiewicz
counting.
We now take a further step and show that RF counting
is likely harder by separating it from $\sharpP$.

\begin{theorem}
[Appendix~\ref{app:sat-gap}]
\label{thm:rf-not-sharpp}
Counting RF-equivalence classes of maximal executions of
bounded deterministic SC read/write programs does not
belong to $\sharpP$ unless
$\mathrm{NP}\subseteq\mathrm{SPP}$.
\end{theorem}

\pheading{Unbiased estimation.}
For choosing a checker, however, exact values may not
be necessary: perhaps approximation or estimation would
suffice.
Although exact RF counting is hard, can we construct
an unbiased estimator whose per-trial running time is
polynomial, as we can for Mazurkiewicz counting?

Recall that Mazurkiewicz counting admits an unbiased
estimator with polynomial time per trial despite being
$\sharpP$-complete.
Unfortunately, we show that RF counting admits no such
estimator when the number of threads is part of the input,
unless $\mathrm{PP}^{\mathrm{NP}}=\mathrm{PP}$.
The same impossibility holds for estimating the reduction
ratio directly.

Formally, an unbiased estimator for a function $f$
is a probabilistic algorithm $A$ satisfying
\[
  \mathbb{E}[A(x)]=f(x)
\]
for every input $x$.
We use the standard bit model:
algorithms use fair random bits and output rational
numbers whose signed numerators and positive denominators
are explicitly represented in binary.
An estimator is polynomial-time if every trial terminates
within a fixed polynomial in the input size,
regardless of its random choices.
This representation permits exponentially large
sampling weights with polynomially many bits;
unbiasedness imposes no bound on their variance.

\begin{theorem}[Appendix~\ref{app:rf-unbiased-barrier}]
\label{thm:rf-unbiased-barrier}
Assume that $\mathrm{PP}^{\mathrm{NP}}\neq\mathrm{PP}$.
Counting RF-equivalence classes of maximal executions
of bounded deterministic SC read/write programs admits
no oracle-free polynomial-time unbiased estimator.
\end{theorem}
A direct consequence is that estimating the reduction ratio directly cannot avoid
the same barrier.

\begin{corollary}[Appendix~\ref{app:ratio-unbiased-barrier}]
\label{cor:ratio-unbiased-barrier}
Assume that $\mathrm{PP}^{\mathrm{NP}}\neq\mathrm{PP}$.
The reduction ratio
\[
  \rho(P)=\frac{C_{\RF}(P)}{C_{\Maz}(P)}
\]
admits no oracle-free polynomial-time unbiased estimator.
\end{corollary}

Given that polynomial-time unbiased estimation is out of
reach, one may ask whether allowing relative error and
a small probability of failure would help.
We show that the ratio admits no FPRAS either, unless
$\mathrm{NP}=\mathrm{RP}$, even with only two threads.

\begin{proposition}[Appendix~\ref{app:ratio-no-fpras}]
Unless $\mathrm{NP}=\mathrm{RP}$, there is no FPRAS
for the reduction ratio $\rho(P)$, even for loop-free
deterministic SC read/write programs with two threads
and Boolean shared variables.
\end{proposition}

\section{Canonical Sampling and Statistical Test}
\label{counting_algo}
The previous section shows that RF counting and computing the reduction ratio are difficult and  unlikely
to admit even an unbiased estimator with polynomial
time per trial when the number of threads $t$ is part
of the input.
But what if we are willing to accept a higher
computational cost in $t$? 
A program may have only a few threads even when
it has many shared locations and a large execution
bound, making this a useful tradeoff to consider.
Theoretically, we can indeed show that, for every fixed $t$, RF counting belongs
to $\sharpP$.
Since every $\sharpP$ function admits a polynomial-time
unbiased estimator, we can recover the same per-trial
guarantee as for Mazurkiewicz counting when $t$ is fixed.

Fixing the number of threads does not, however, make
exact counting or approximation easy.
Even with only two threads, RF and Mazurkiewicz counting
remain $\sharpP$-complete, and computing the reduction
ratio remains $\sharpP$-hard.
None of these problems admits an FPRAS unless
$\mathrm{NP}=\mathrm{RP}$.
Thus, unbiased estimation remains a useful target,
just as it is for Mazurkiewicz counting.

In this section, we first establish $\sharpP$ membership by constructing
a canonicality checker that recognizes exactly one
Mazurkiewicz representative of each RF class in
polynomial time for fixed $t$. Note that when $t$ is not fixed, this is a $\mathrm{NP}$ solver.

We then use this checker to construct our unbiased
RF-count estimator.
Each trial performs polynomial-time computation apart
from one call to a witness-producing SC consistency
checker, whose cost can grow rapidly with $t$.
Combining the RF and Mazurkiewicz estimates gives
a strongly consistent estimator of the reduction ratio.
Finally, we present a statistical test based on
this ratio estimator to support checker selection
by testing whether the ratio lies below or above
a given threshold.

\pheading{A canonical representation function.}
The key insight behind our algorithm is that SC
consistency checking can be used to select a canonical
representative for each RF class.
The SC consistency procedure of
Abdulla et al.~\cite{rfsc2019} searches for an execution
realizing a given RF graph.
The write orders in this execution determine a compatible
coherence order.
By fixing a deterministic search policy that depends
only on the RF graph, we can use the first completion
found as its canonical representative.

\begin{proposition}[Canonical completion: Appendix~\ref{app:canonical-completion}]
\label{prop:canonical-completion}
For every fixed number of threads $t$, there is a
deterministic polynomial-time function $\kappa$ that,
given an RF graph $H$ on at most $t$ threads, returns
an SC-consistent full graph $\kappa(H)$ satisfying
\[
  \pi_{\RF}(\kappa(H))=H,
\]
or returns $\bot$ if no such completion exists.
\end{proposition}

With this function, we can count RF classes by guessing
their RF graphs and checking whether they have a
completion.

\begin{proposition}[RF counting with a fixed thread bound: Appendix~\ref{app:rf-fixed-t-sharpp}]
\label{prop:rf-fixed-t-sharpp}
For every fixed $t$, the RF-counting problem of
Section~\ref{sec:complexity}, restricted to programs
with at most $t$ threads, belongs to $\sharpP$.
It therefore admits an unbiased estimator with
polynomial time per trial.
\end{proposition}

\pheading{Estimating the RF count and reduction ratio.}
The result above shows that, for every fixed $t$,
An unbiased RF-count estimator
with polynomial time per trial must exist.
This theoretical insight guides our algorithm design,
but such an algorithm may still have such high variance that it fails
to produce useful estimates within a reasonable
time budget. 

How should we sample to obtain useful estimates
in practice?
Balasubramanian et al.~\cite{pldi26} address this
question for Mazurkiewicz counting.
Their approach is to use TruSt's exploration tree
as the sampling space.
This tree is organized by the program's semantics:
its successful paths generate maximal execution graphs,
with exactly one path for each Mazurkiewicz class.
The branching choices along a sampled path determine
its sampling probability, allowing the estimator
to compute the corresponding weight during sampling.  

Our idea is to build on this design.
With a small modification using our canonicality
checker, we can reuse these weighted execution
samples to obtain an unbiased RF-count estimator. Let us now refer to the Mazurkiewicz estimator of
Balasubramanian et al.~\cite{pldi26} as $\mazsampler$.
Each trial of $\mazsampler$ samples a path through
TruSt's exploration tree in polynomial time.
\footnote{
For clarity, we describe the single-path version.
Their implementation uses stochastic enumeration
to advance multiple sampled paths together and
reduce variance.
Our construction extends to this version with
the same unbiasedness and strong consistency
guarantees, as shown in the appendix.}
If the path reaches a maximal execution graph $G$,
$\mazsampler$ returns $(G,w)$, where $w=1/q(G)$ and
$q(G)$ is the probability of sampling $G$.
Otherwise, it returns $(\bot,0)$.
With this weighting, each Mazurkiewicz class
contributes one in expectation.

A natural idea is to collect $N$ maximal execution samples
from $\mazsampler$, check which are canonical, and use
the fraction of canonical samples to estimate $\rho$.
If the samples were uniform over maximal Mazurkiewicz
classes, this would give an unbiased estimate: exactly
one class per RF class passes the canonicality check.
However, $\mazsampler$ does not sample uniformly, so
the canonical-hit rate can differ from the reduction ratio.

\begin{example}
\label{sample-example}
Consider again the program in
Figure~\ref{fig:prelim-rf-projection}, which has three
maximal Mazurkiewicz classes and two RF classes.
The graphs $G_a$ and $G_b$ shown in the figure belong
to the same RF class.
The third graph, denoted by $G_c$, has $R$ reading
from $W_1$ and belongs to the other RF class.
Suppose that the canonicality checker selects
$G_a$ and $G_c$ as representatives, and that
$\mazsampler$ samples $G_a$, $G_b$, and $G_c$
with probabilities $1/2$, $1/4$, and $1/4$, respectively.
The expected canonical-hit rate is then
$1/2+1/4=3/4$, whereas the reduction ratio is $\rho=2/3$.
Increasing the number of samples does not remove
this bias.
\end{example}

Nor can we simply replace $\mazsampler$ with an
efficient uniform sampler.
The construction used to prove
Proposition~\ref{prop:ratio-no-fpras} also rules out
polynomial-time uniform sampling of maximal
Mazurkiewicz classes unless $\mathrm{NP}=\mathrm{RP}$,
even for two-thread programs.

However, we can account for the unequal sampling
probabilities using the weights already returned by
$\mazsampler$.
With these weights, each Mazurkiewicz class contributes
one in expectation.
For RF counting, we keep the weight only when the
sampled graph is canonical.
Exactly one Mazurkiewicz class then contributes to
each RF class, giving an unbiased RF-count estimate.

As shown in Algorithm~\ref{alg:canonical-rejection},
our procedure $\pairedestimation$ takes the sampled
execution graph $G$ and its weight $w$, and runs
the canonicality checker on $G$.
It keeps the original weight as the Mazurkiewicz
estimate $X=w$, and sets the RF estimate to $Y=w$
if $G$ is canonical and $Y=0$ otherwise. Note, however, that we cannot take the ratio for each
pair and then average these ratios.
For $X>0$, $Y/X$ is simply one if the sample is
canonical and zero otherwise, so this would give
the same canonical-hit estimator discussed in
the example above.

This procedure is implemented by \name{} (Coarsening Estimator) in
Algorithm~\ref{alg:canonical-rejection}.
It runs $\pairedestimation$ independently
$N$ times and returns the ratio of the accumulated
RF and Mazurkiewicz contributions:
\[
  \widehat{\rho}_N
  =
  \frac{\sum_{i=1}^{N}Y_i}
       {\sum_{i=1}^{N}X_i}.
\]

\begin{proposition}
\label{prop:paired-estimation}
For every program $P$, the pair $(X,Y)$ returned
by $\pairedestimation(P)$ satisfies
\[
  \mathbb{E}[X]=C_{\Maz}(P),
  \qquad
  \mathbb{E}[Y]=C_{\RF}(P).
\]
Using independent trials, $\name(P,N)$
gives a strongly consistent estimate of the reduction ratio:
\[
  \widehat{\rho}_N
  \xrightarrow[N\to\infty]{\mathrm{a.s.}}
  \rho(P).
\]
\end{proposition}

\begin{example}
Consider the same program and sampling probabilities
as in Example~\ref{sample-example}.
When $G_a$ is sampled, \name{} records $(X,Y)=(2,2)$.
When $G_b$ is sampled, it records $(X,Y)=(4,0)$,
since this graph is not canonical.
When $G_c$ is sampled, it records $(X,Y)=(4,4)$.
Thus,
\[
\begin{aligned}
  \mathbb{E}[X]
  &= \frac12\cdot2+\frac14\cdot4+\frac14\cdot4=3,\\
  \mathbb{E}[Y]
  &= \frac12\cdot2+\frac14\cdot0+\frac14\cdot4=2.
\end{aligned}
\]
Both count estimates are unbiased, and the ratio of
their accumulated contributions converges almost
surely to the true ratio $\rho=2/3$.
\end{example}

\begin{algorithm}[t]
\caption{\name{}: canonical-rejection estimation}
\label{alg:canonical-rejection}
\begin{algorithmic}[1]
\Procedure{\pairedestimation}{$P$}
    \State $(G,w) \gets \Call{\mazsampler}{P}$
    \If{\Call{\iscanonical}{$G$}}
        \State \Return $(w,w)$
    \EndIf
    \State \Return $(w,0)$
\EndProcedure
\Statex
\Procedure{\name}{$P,N$}
    \State $S_{\Maz}\gets 0$; $S_{\RF}\gets 0$
    \For{$i\gets 1$ to $N$}
        \State $(X,Y)\gets\Call{\pairedestimation}{P}$
        \State $S_{\Maz}\gets S_{\Maz}+X$;
               $S_{\RF}\gets S_{\RF}+Y$
    \EndFor
    \State \Return $S_{\RF}/S_{\Maz}$
\EndProcedure
\end{algorithmic}
\end{algorithm}

\pheading{A statistical test for checker selection.}
As discussed in Section~\ref{sec:motivation},
let $h$ denote the average cost of exploring an RF class
relative to a Mazurkiewicz class.
We predict RF exploration to be faster when
$\rho<\eta:=1/h$, and Mazurkiewicz exploration
to be faster when $\rho>\eta$.

We therefore ask whether our samples provide
sufficient evidence for either conclusion.
We use a standard $Z$-score test to assess whether
the estimated ratio is sufficiently far from
the threshold relative to its sampling uncertainty.
If the evidence is insufficient, we report
the result as inconclusive.

Fix $\eta$ before collecting the estimation samples.
We collect a predetermined number $N$ of independent
paired estimates and define
\[
  D_i=Y_i-\eta X_i.
\]
Then
\[
  \mathbb{E}[D_i]
  =
  C_{\Maz}(P)(\rho-\eta).
\]
Thus, testing whether $\rho$ is below or above $\eta$
amounts to testing whether the mean of $D_i$ is
negative or positive.

Let $\overline D_N$ and $s_D$ be the sample mean and
sample standard deviation of these differences.
We use two one-sided normal tests, each with
asymptotic significance level $0.05$.
The test reports the RF direction when
\[
  \overline D_N<-1.645\,\frac{s_D}{\sqrt N},
\]
and the Mazurkiewicz direction when
\[
  \overline D_N>1.645\,\frac{s_D}{\sqrt N}.
\]
Otherwise, the result is inconclusive.

We also study finite-sample accuracy in
Appendix~\ref{app:estimation-variance}.
Our bound accounts for two sources of error:
substantial underestimation of the Mazurkiewicz count,
and errors in the two count estimates that do not
cancel when we take their ratio.
Even inaccurate count estimates can give an accurate
ratio when they err by similar proportions.

\pheading{An alternative RF estimator.}
We end this section with a natural question:
can we sample RF classes directly, instead of sampling
Mazurkiewicz classes and rejecting non-canonical graphs?
In principle, an RF exploration tree could support
the same estimation approach: each sampled maximal
RF graph would contribute its inverse sampling probability if the tree is also branch independent.

TruSt/wb~\cite{TruSt2022} offers a possible route
to such an estimator.
The original paper describes an RF variant of TruSt
and sketches its adaptation to SC,
but does not provide a corresponding correctness proof
for the SC adaptation.
Developing this adaptation into an estimator would
require establishing its correctness and the properties
needed for efficient sampling and out of the scope of this work.

Nidhugg/rfsc~\cite{rfsc2019} also explores RF classes
under SC, but its pending exploration choices are
updated using information obtained from other branches.
Unlike TruSt, it therefore does not directly support
sampling a single path independently of the rest
of the exploration.
How to obtain an efficient estimator from this
algorithm remains unclear.

\section{Experimental Evaluation}
\label{sec:experiments}

In this section, we gauge \name{} to choose between RF and Mazurkiewicz checkers
and ask whether this choice saves enough verification time
to pay for the estimation and checker pilots.
Each estimation run has a one-minute budget. We combine
its ratio estimate with the exploration rates measured during
the pilots.
We first examine the accuracy of the estimates and the choices
they produce.
We then compare the total cost with two alternatives: 
(a) always using
one checker and (b) running both in round robin.
To understand the role of sampling weights and population size,
we compare 3 population budgets with a baseline that uses
the same samples without their weights.

\subsection{Experimental Setup and Benchmarks}
\label{sec:exp-setup}

\pheading{Implementation and platform.}
We used OpenAI's Codex (GPT 6 Astra) and Anthropic's Claude Code (Fable 5.2) to implement \name{} in GenMC~0.19.0.
We compare GenMC in Mazurkiewicz mode with Nidhugg in
RF-SC mode at revision \texttt{845a1463}.
The implementation is built with GCC~14.2.0 and LLVM~19.1.7
in \texttt{RelWithDebInfo} mode with \texttt{-O2} optimization.
Each measurement runs in its own aarch64 Linux Docker
container with one CPU and 16\,GiB of memory, on a shared
Apple M3 Max host with 14 CPU cores and 36\,GiB of memory. 

\pheading{Benchmarks.}
Our suite contains 34 configurations of bounded programs
from 24 families,
covering concurrent data structures, synchronization algorithms,
a simplified device-driver model, and synthetic workloads.
We draw or adapt programs from the GenMC and Nidhugg
suites~\cite{genmc-benchmarks,nidhugg-benchmarks}, and include
the optimistic-list algorithm of Herlihy et al.~\cite{book:herlihy-shavit}
and synthetic benchmarks.
Table~\ref{tab:estimates-and-choices} lists the configurations,
which vary thread counts, operation counts, data-structure
sizes, and loop bounds.

Thirteen configurations have $\rho=1$, each from
a different family.
We also include one configuration each of Write-only
and Control-flow, with RF counts 1 and 531, respectively.
For the remaining nine families, the reduction can change
with instance size, so we include several sizes where possible.
For example, the $\rho$ values are 0.020 and 0.011
on two sizes of Sigma, and 0.26 and 0.21 on two sizes of QU.
We include three sizes each of Eratosthenes and Sigma, and
two each of ApproxDS, Scull, RW-SC, Opt-lock, QU, and
Race-parametric.
We use only one size of Lamport: the next smaller instance
has just 16 executions, while neither checker completes
the next larger one within three hours.

\pheading{Canonicality checking.}
We implement \iscanonical{} using GenMC's consistency
validator with deterministic backtracking over coherence
orders and constraint propagation
inspired by Abdulla et al.~\cite{rfsc2019}.
The search depends only on the input RF graph.
We also implemented their polynomial-time procedure for
fixed $t$.
Although it has a better worst-case bound, backtracking
was slightly faster in our preliminary experiments,
so we use it here.

\pheading{Sampling protocol.}
Similar to \textsc{TeStor}~\cite{pldi26}
\name{} uses the stochastic-enumeration\cite{vaisman2017} to reduce variance.
The population budget $b$ limits the number of sampled
paths advanced together within a trial.
We evaluate $b\in\{1,20,50\}$; with $b=1$, each trial
follows a single random path.
We repeat trials for one minute, allowing the active trial
to finish when the budget expires.
Warmup made little difference in our preliminary runs,
so we omit it from the reported experiments.
Each completed trial returns paired contributions $(X_i,Y_i)$.
After $n$ trials, we compute
\[
 \widehat C_{\Maz}=\frac{1}{n}\sum_{i=1}^{n}X_i,
 \qquad
 \widehat C_{\RF}=\frac{1}{n}\sum_{i=1}^{n}Y_i,
 \qquad
 \widehat\rho=\frac{\sum_{i=1}^{n}Y_i}{\sum_{i=1}^{n}X_i}.
\]
For each configuration and population budget, we repeat
the experiment with ten random seeds.
Each run receives the full sampling budget and produces
its own estimates and choice of checker.
Across the 34 configurations, the median run completes
about 60{,}000 trials with $b=1$, 1{,}800 with $b=20$,
and 640 with $b=50$.

\pheading{Baseline: Canonical hit.}
We implement the generally biased estimator discussed
in Example~\ref{sample-example} as a baseline,
which we call \emph{Canonical hit} (CH).
It uses exactly the same single-path samples as \name{}
with $b=1$, and returns the fraction of sampled normally
terminating executions that are canonical.
We compare \name{} with CH to see whether this bias
hurts estimation accuracy in practice.

\pheading{Blocked executions.}
A checker may encounter a failed \texttt{assume} during
exploration, including one introduced by a spinloop transformation.
Checkers such as GenMC and Nidhugg report these runs as
\emph{blocked executions} and exclude them from their
normal-execution counts.
Which blocked executions are explored depends on the
checker's program transformations, treatment of bounds,
and exploration algorithm.
For example, Awamoche~\cite{awamoche} can avoid many blocked
executions arising from await and confirmation-CAS loops.
We count classes of normally terminating executions under
the chosen equivalence.
Blocked samples contribute zero to both the RF and
Mazurkiewicz estimates.

\subsection{Choosing Between RF and Mazurkiewicz Checkers}
\label{sec:exp-selection}

As discussed in Section~\ref{sec:motivation}, the reduction
in execution classes must offset the additional cost of
exploring each RF class.
We use two maintained, state-of-the-art DPOR model checkers:
GenMC in Mazurkiewicz mode (\emph{Maz}) and
Nidhugg in RF-SC mode (\emph{RF}).
Choosing tools with different exploration algorithms
and implementations lets us test whether our estimates
of the reduction between equivalence relations can
help select the faster checker across different
DPOR implementations.

\pheading{Predicting which checker will finish sooner.}
A short pilot tells us how quickly each checker explores;
the ratio estimate tells us how much exploration RF can avoid.
We run each checker for up to one minute to obtain rates
$v_M$ and $v_R$. We compute each pilot rate by dividing the checker's
reported normal-execution count by its running
time.

These rates estimate the relative per-class cost $h$
as $H_{\mathrm{pilot}}=v_M/v_R$.
The corresponding threshold is
$\eta=1/H_{\mathrm{pilot}}=v_R/v_M$.
We predict RF checker to be faster when $\widehat\rho<\eta$ and Mazurkiewicz otherwise.

\pheading{Full checker runs and exact counts.}
We use the estimators to select the largest workload
for which we expect at least one checker to finish
within three hours.
We then run each checker with a three-hour timeout
to measure its verification time and determine
which checker is faster.

To evaluate estimation accuracy, we obtain both exact counts
using GenMC, so they follow the same bounded-program semantics
as our estimator.
We count canonical representatives rather than using
Nidhugg's RF count because the two tools can transform
bounded programs differently.
In particular, an execution that reaches a bound before
terminating may be counted as complete by one tool and
blocked by the other.
Nidhugg's reported RF count agrees with GenMC's canonical
count on 22 of the 26 configurations on which both
checkers complete.
On the remaining four, Nidhugg reports approximately
8\% to 21\% fewer executions.

For each configuration on which GenMC completes its
Mazurkiewicz exploration, the number of explored classes
gives $C_{\Maz}$, and the number passing the canonicality
check gives $C_{\RF}$.
This yields the exact ratio $\rho=C_{\RF}/C_{\Maz}$
for 29 configurations.
The remaining five are those on which Maz times out for which we don't have a ground truth of $\rho$.
\subsection{Results}
\label{sec:exp-results}

\begin{table*}[t]
\centering
\tiny
\setlength{\tabcolsep}{3pt}
\renewcommand{\arraystretch}{1.1}
\resizebox{\textwidth}{!}{%
\begin{tabular}{lrrrrrcrcrcrc}
\toprule
Configuration & $\rho$ & $\eta$ & $T_M$ (s) & $T_R$ (s)
& \multicolumn{2}{c}{Canonical hit}
& \multicolumn{2}{c}{\name{}, $b{=}1$}
& \multicolumn{2}{c}{\name{}, $b{=}20$}
& \multicolumn{2}{c}{\name{}, $b{=}50$} \\
\cmidrule(lr){6-7}\cmidrule(lr){8-9}\cmidrule(lr){10-11}\cmidrule(lr){12-13}
& & & & & q-err. & correct & q-err. & correct & q-err. & correct & q-err. & correct \\
\midrule
ApproxDS (N6, M4) & $0.0202$ & $0.0961$ & $476.72$ & $93.95$ & $4.86$ & 0/10 & $1.45$ & 10/10 & $1.23$ & 10/10 & $1.10$ & 10/10 \\
Barrier (L2) & $1$ & $0.295$ & $553.65$ & $1322.15$ & $1$ & 10/10 & $1$ & 10/10 & $1$ & 10/10 & $1$ & 10/10 \\
DQ (L3) & $1$ & $0.423$ & $827.05$ & $1960.58$ & $1$ & 10/10 & $1$ & 10/10 & $1$ & 10/10 & $1$ & 10/10 \\
Eratosthenes SC (N14, U16) & $0.00757$ & $0.127$ & $435.98$ & $26.60$ & $61.3$ & 0/10 & $3.30$ & 9/10 & $1.21$ & 10/10 & $1.11$ & 10/10 \\
Lamport (N3, U16) & $0.428$ & $0.271$ & $1.01$ & $1.70$ & $1.18$ & 10/10 & $1.43$ & 9/10 & $1.01$ & 10/10 & $1.01$ & 10/10 \\
MCS (N4, U8) & $1$ & $0.245$ & $3080.46$ & $>9297.32$ (OOM) & $1$ & 10/10 & $1$ & 10/10 & $1$ & 10/10 & $1$ & 10/10 \\
MS queue (W3, RW1) & $1$ & $0.470$ & $6350.25$ & $>10800$ & $1$ & 10/10 & $1$ & 10/10 & $1$ & 10/10 & $1$ & 10/10 \\
Mutex Musl (L1) & $1$ & $0.471$ & $149.23$ & $249.23$ & $1$ & 10/10 & $1$ & 10/10 & $1$ & 10/10 & $1$ & 10/10 \\
Optimistic list (L2) & $1$ & $0.193$ & $125.52$ & $719.98$ & $1$ & 10/10 & $1$ & 10/10 & $1$ & 10/10 & $1$ & 10/10 \\
QU (4,3,3) & $0.209$ & $0.178$ & $1596.26$ & $1910.45$ & $4.25$ & 10/10 & $2.54$ & 4/10 & $1.14$ & 9/10 & $1.10$ & 9/10 \\
RW-SC (N7) & $0.0103$ & $0.178$ & $680.81$ & $41.81$ & $2.57$ & 10/10 & $2.27$ & 10/10 & $1.65$ & 10/10 & $1.47$ & 10/10 \\
Scull (R3, W2) & $0.187$ & $0.143$ & $2090.16$ & $2992.09$ & $2.85$ & 10/10 & $1.92$ & 5/10 & $1.62$ & 6/10 & $1.16$ & 10/10 \\
Spinlock (N4, U6) & $1$ & $0.316$ & $3194.93$ & $9332.99$ & $1$ & 10/10 & $1$ & 10/10 & $1$ & 10/10 & $1$ & 10/10 \\
STC (P5, 2/2/2/2, U16) & $1$ & $0.208$ & $764.46$ & $3705.85$ & $1$ & 10/10 & $1$ & 10/10 & $1$ & 10/10 & $1$ & 10/10 \\
Ticketlock (L1) & $1$ & $0.384$ & $2734.27$ & $7636.23$ & $1$ & 10/10 & $1$ & 10/10 & $1$ & 10/10 & $1$ & 10/10 \\
Treiber dynamic (W4, RW1) & $1$ & $0.480$ & $2946.04$ & $8215.90$ & $1$ & 10/10 & $1$ & 10/10 & $1$ & 10/10 & $1$ & 10/10 \\
Treiber static (L3) & $1$ & $0.147$ & $353.01$ & $2749.56$ & $1$ & 10/10 & $1$ & 10/10 & $1$ & 10/10 & $1$ & 10/10 \\
TTAS & $1$ & $0.351$ & $64.80$ & $170.55$ & $1$ & 10/10 & $1$ & 10/10 & $1$ & 10/10 & $1$ & 10/10 \\
Write-only (N11) & $2.51\!\times\!10^{-8}$ & $0.00233$ & $1572.75$ & $0.0102$ & $\infty$ & 10/10 & $\infty$ & 10/10 & $\infty$ & 10/10 & $\infty$ & 10/10 \\
TWA (N4, A4, U16) & $1$ & $0.528$ & $9883.11$ & $>10800$ & $1$ & 10/10 & $1$ & 10/10 & $1$ & 10/10 & $1$ & 10/10 \\
ApproxDS (N7, M5) & -- & $0.0890$ & $>10800$ & $2637.40$ & -- & 10/10 & -- & 10/10 & -- & 10/10 & -- & 10/10 \\
Eratosthenes SC (N13, U16) & $0.0148$ & $0.134$ & $124.76$ & $14.24$ & $30.1$ & 0/10 & $1.71$ & 10/10 & $1.07$ & 10/10 & $1.10$ & 10/10 \\
Scull (R2, W2) & $0.208$ & $0.135$ & $152.63$ & $230.64$ & $2.65$ & 10/10 & $1.32$ & 9/10 & $3.23$ & 8/10 & $1.31$ & 9/10 \\
RW-SC (N8) & -- & $0.160$ & $>10800$ & $1026.85$ & -- & 10/10 & -- & 10/10 & -- & 10/10 & -- & 10/10 \\
Sigma (N9) & $0.0105$ & $0.155$ & $1422.42$ & $97.57$ & $4.65$ & 10/10 & $1.05$ & 10/10 & $1.04$ & 10/10 & $1.04$ & 10/10 \\
Opt-lock (N13) & $7.34\!\times\!10^{-5}$ & $0.113$ & $86.35$ & $0.0766$ & $7570$ & 0/10 & $5.94$ & 10/10 & $1.64$ & 10/10 & $1.49$ & 10/10 \\
Control-flow (N9) & $2.76\!\times\!10^{-6}$ & $0.160$ & $8856.33$ & $0.2057$ & $\infty$ & 10/10 & $\infty$ & 10/10 & $\infty$ & 10/10 & $\infty$ & 10/10 \\
Race-parametric (N7) & $0.153$ & $0.0563$ & $403.64$ & $729.21$ & $3.74$ & 10/10 & $2.09$ & 9/10 & $1.12$ & 10/10 & $1.05$ & 10/10 \\
QU (3,3,3) & $0.264$ & $0.185$ & $110.41$ & $155.00$ & $3.45$ & 10/10 & $1.75$ & 8/10 & $1.06$ & 10/10 & $1.05$ & 10/10 \\
Sigma (N8) & $0.0199$ & $0.158$ & $73.92$ & $9.69$ & $3.74$ & 10/10 & $1.03$ & 10/10 & $1.03$ & 10/10 & $1.03$ & 10/10 \\
Race-parametric (N6) & $0.198$ & $0.0826$ & $36.58$ & $84.60$ & $3.01$ & 10/10 & $1.45$ & 9/10 & $1.05$ & 10/10 & $1.04$ & 10/10 \\
Sigma (N10) & -- & $0.151$ & $>10800$ & $1097.66$ & -- & 10/10 & -- & 10/10 & -- & 10/10 & -- & 10/10 \\
Opt-lock (N512) & -- & $0.190$ & $>10800$ & $181.86$ & -- & 0/10 & -- & 10/10 & -- & 10/10$^\dagger$ & -- & 10/10$^\dagger$ \\
Eratosthenes SC (N17, U32) & -- & $0.115$ & $>10800$ & $241.55$ & -- & 0/10 & -- & 10/10 & -- & 10/10 & -- & 10/10 \\
\midrule
Mean q-error (27 configurations) &  &  &  &  & $286$ &  & $1.56$ &  & $1.19$ &  & $1.08$ &  \\
Mean q-error, $\rho<1$ (14 configurations) &  &  &  &  & $550$ &  & $2.09$ &  & $1.36$ &  & $1.15$ &  \\
Correct choices &  &  &  &  &  & 280/340 &  & 322/340 &  & 333/340 &  & 338/340 \\
\bottomrule
\end{tabular}%
}
\caption{Ratio estimates and checker choices with a one-minute
sampling budget.
$\rho$: exact RF/Maz count ratio;
$\eta$: RF/Maz pilot rate ratio;
$T_M$, $T_R$: Maz and RF verification times.
For each method, \emph{q-err.} gives the mean q-error over
ten seeds, and \emph{correct} counts choices of the faster checker.
Mean q-errors at the bottom exclude Write-only and Control-flow,
whose zero estimates give infinite q-error.
Missing exact ratios are marked --; $>10800$ denotes a timeout,
and (OOM) denotes memory exhaustion.
$^\dagger$A single trial exceeds the sampling budget;
the run ends when that trial completes.}
\label{tab:estimates-and-choices}
\end{table*}

\pheading{How accurate are the estimates?}
We measure ratio accuracy using q-error,
\[
 q(\hat r,r)=\max\{\hat r/r,r/\hat r\}.
\]
A q-error of one is exact, and two means that the estimate
differs from the reference by a factor of two in either
direction.
Table~\ref{tab:estimates-and-choices} reports the arithmetic
mean of the ten seeds' q-errors for each configuration.
A zero estimate against a positive reference has infinite
q-error.
This occurs on Write-only and Control-flow, where canonical
executions are rarely sampled.
Their mean q-errors are therefore infinite.
We examine these two cases below and compute the mean
q-error across the other 27 configurations with reference ratios.

On these 27 configurations, \name{} has a mean q-error
of 1.56 with $b=1$, 1.19 with $b=20$, and 1.08 with $b=50$.
The ratio estimate is exact for every seed and population budget
on the 13 configurations with $\rho=1$.
Every Mazurkiewicz class is then canonical, so the RF
and Mazurkiewicz contributions agree on every trial.
On the remaining 14 configurations, the mean q-errors
are 2.09, 1.36, and 1.15, respectively.
A larger population lets each trial cover more of the tree
at the cost of completing fewer trials.
Here, the additional work within each trial pays off in
lower mean q-error.

CH has a mean q-error of 286 over the same 27 configurations,
and 550 over the 14 with $\rho<1$.
It overestimates $\rho$ on 13 of these 14 configurations,
by factors between 2.6 and 7{,}570, and underestimates it
only on Lamport.
The weights correct for these unequal probabilities.
Applying them to the same samples (\name{} with b=1) reduces the mean q-error
over the 27 configurations from 286 to 1.56.

On Write-only and Control-flow, the real $\rho$ values are
extremely small, and most runs sample no canonical
representative, yielding zero estimates and infinite q-error.
Every estimate nevertheless falls below the pilot threshold,
so all runs correctly choose RF.

\pheading{How often do we choose the faster checker?}
RF finishes sooner on 14 of the 34 configurations, and
Maz on the other 20.
Always choosing Maz is therefore correct in 200 of
340 runs (58.8\%), and always choosing RF in 140 (41.2\%).
Using \name{}'s estimates and the pilot rates, we choose
the faster checker in 322 runs (94.7\%) with $b=1$,
333 (97.9\%) with $b=20$, and 338 (99.4\%) with $b=50$.
With $b=50$, all ten seeds choose correctly on 32 of
the 34 configurations.
The two incorrect choices are one seed each on QU (4,3,3)
and Scull (R2, W2).
CH chooses correctly in 280 runs (82.4\%).

Choosing solely by the pilot exploration rates would
select Maz on every configuration, since $\eta<1$ for every configuration.
This is equivalent to Always Maz, which misses the
fourteen configurations on which RF finishes sooner.
Therefore the ratio estimate helps identify when RF's smaller
exploration space makes up for its lower exploration rate.

\pheading{Does the time saved pay for the choice?}
A correct choice is useful only if it saves more time
than we spend making it.
We compare CH and \name{} at the three population budgets
with \emph{Always Maz}, \emph{Always RF}, and
\emph{Ideal round robin}.
We exclude nine configurations on which at least one
checker finishes during its pilot, since no subsequent
choice is needed.
This leaves 25 configurations and, with ten seeds each,
250 choices for each sampling setting.
RF is faster on seven of these configurations.

Ideal round robin gives both checkers equal processor
shares and stops when either finishes.
With independent memory limits and no switching overhead,
its cost is
\[
 C_{\mathrm{RR}}=2\min(T_M,T_R).
\]
We reconstruct the cost of our procedure from the recorded
runs, charging two one-minute pilots, the measured estimation
time $t_{\mathrm{est}}$, and a fresh run of the selected
checker $S$:
\[
 C_{\mathrm{select}}=120+t_{\mathrm{est}}+T_S.
\]
The measured estimation time includes finishing any trial
still in progress when the budget expires.
We reuse the selected checker's recorded runtime $T_S$
across the ten seeds.
We also report its slowdown, $T_S/\min(T_M,T_R)$,
which measures the cost of the choice without the
preliminary work.
Timeouts and memory exhaustion give lower bounds
on slowdown and total cost.
\begin{table}[t]
\centering
\tiny
\begin{tabular}{lrrrr}
\toprule
Strategy & Correct choices & Mean slowdown & Worst slowdown & Total cost (h) \\
\midrule
Canonical hit & 220/250 & $>1.463$ & $>59.4$ & $>185.61$ \\
\name{}, $b{=}1$ & 235/250 & $1.018$ & $1.81$ & $127.64$ \\
\name{}, $b{=}20$ & 243/250 & $1.010$ & $1.51$ & $127.21$ \\
\name{}, $b{=}50$ & 248/250 & $1.002$ & $1.51$ & $127.05$ \\
\midrule
Always Maz & 180/250 & $>2.074$ & $>59.4$ & $>253.55$ \\
Always RF & 70/250 & $>1.846$ & $>7.79$ & $>217.65$ \\
Ideal round robin & -- & $2.000$ & $2.00$ & $226.43$ \\
\bottomrule
\end{tabular}
\caption{Checker selection on 25 configurations where neither
checker finishes during its one-minute pilot, with ten
decisions per configuration.
Slowdown is relative to the faster checker; we report
the geometric mean and maximum.
Costs for CH and \name{} include both pilots, estimation,
and a fresh checker run.
$>$ denotes a lower bound due to timeout or memory exhaustion.}
\label{tab:selection-strategies}
\end{table}

Table~\ref{tab:selection-strategies} shows that \name{}
with $b=50$ chooses the faster checker in 248 of 250
decisions, with a geometric mean slowdown of 1.002
and a worst slowdown of 1.51.
Its reconstructed total cost is 127.05 hours,
43.9\% below ideal round robin's 226.43 hours,
at least 49.9\% below the Always Maz bound of 253.55 hours,
and at least 41.6\% below the Always RF bound of 217.65 hours.
The other population budgets give similar total costs,
despite making more incorrect choices:
127.21 hours with 243 correct decisions for $b=20$,
and 127.64 hours with 235 correct decisions for $b=1$.

CH chooses correctly in 220 of the 250 decisions.
Its 30 incorrect choices occur on ApproxDS (N6, M4),
Opt-lock (N512), and Eratosthenes SC (N17, U32).
It chooses Maz on these programs, although RF verifies
them much faster.
Its total cost is at least 185.61 hours.
We note that both methods benefit from the 13 configurations with
$\rho=1$, on which every seed correctly chooses Maz.
On the remaining 12 configurations (120 decisions),
\name{} with $b=50$ costs 34.36 hours,
36.4\% less than ideal round robin's 54.06 hours,
whereas CH costs at least 92.92 hours.

In all but twenty runs of \name{}, every estimation trial is fast and the last trial extends
the one-minute budget by less than a second. The exception is Opt-lock (N512), where a single trial
can cost more than verifying the program with RF.
With $b=20$ and $b=50$, the first trial is still running
when the minute expires and finishes after about 190
and 500 seconds respectively.
The resulting total costs are about 490 and 800 seconds,
compared with 364 seconds for ideal round robin.
Nidhugg verifies this program in 182 seconds by exploring
6{,}139 executions, while a single trial with $b=20$
expands about 30{,}000 nodes of the Mazurkiewicz tree.
The smaller population avoids this expensive trial:
with $b=1$, a one-minute run completes about 507{,}000
trials, and the total cost is 363 seconds. This illustrates the limitation of \name{} with large budgets: Here, RF reduces the exploration space so substantially
that advancing many Mazurkiewicz paths within a single
trial costs more than completing RF verification.
The larger population is unnecessary for this choice:
with $b=1$, \name{} already selects RF correctly in
all ten runs, at a much lower estimation cost.

\pheading{When is the choice difficult?}
When $\rho$ is close to the pilot threshold $\eta$,
a small estimation error can change the choice.
Table~\ref{tab:near-threshold} lists the seven configurations
on which Maz is faster and \name{} makes at least one
incorrect choice across the tested seeds and population budgets.

The closer $\rho/\eta$ is to one, the smaller the
underestimate needed to select RF incorrectly.
All incorrect choices with $b\ge20$ occur in this group.

Among the configurations requiring a choice after the
pilots, these mistakes incur only modest overhead.
As Table~\ref{tab:selection-strategies} shows, even with
$b=1$, \name{} has a geometric mean slowdown of 1.018
and a worst slowdown of 1.81.
CH's worst slowdown exceeds 59.4, because its bias can
hide a substantial advantage of RF.

\begin{table}[t]
\centering
\tiny
\begin{tabular}{lrrrrrrr}
\toprule
& & & & & \multicolumn{3}{c}{Incorrect choices (of 10)} \\
\cmidrule(lr){6-8}
Configuration & $\rho$ & $\eta$ & $\rho/\eta$ & $T_R/T_M$ & $b{=}1$ & $b{=}20$ & $b{=}50$ \\
\midrule
QU (4,3,3) & $0.209$ & $0.178$ & $1.18$ & $1.20$ & 6 & 1 & 1 \\
Scull (R3, W2) & $0.187$ & $0.143$ & $1.31$ & $1.43$ & 5 & 4 & 0 \\
QU (3,3,3) & $0.264$ & $0.185$ & $1.43$ & $1.40$ & 2 & 0 & 0 \\
Scull (R2, W2) & $0.208$ & $0.135$ & $1.54$ & $1.51$ & 1 & 2 & 1 \\
Lamport (N3, U16) & $0.428$ & $0.271$ & $1.58$ & $1.69$ & 1 & 0 & 0 \\
Race-parametric (N6) & $0.198$ & $0.0826$ & $2.40$ & $2.31$ & 1 & 0 & 0 \\
Race-parametric (N7) & $0.153$ & $0.0563$ & $2.72$ & $1.81$ & 1 & 0 & 0 \\
\bottomrule
\end{tabular}
\caption{The configurations with $\rho<1$ on \name{} failed to predict with $100\%$, ordered by $\rho/\eta$.}
\label{tab:near-threshold}
\end{table}

\pheading{Do short pilots capture relative exploration costs?}
We compare the RF/Maz cost per reported execution during
the pilots and over the complete runs:
\[
 H_{\mathrm{pilot}}=\frac{1}{\eta},
 \qquad
 H_{\mathrm{full}}=\frac{T_R/N_R}{T_M/N_M},
\]
where $N_M$ and $N_R$ are the respective normal-execution
counts from the complete runs.
A value above one means that RF spends more time per
reported execution.

To measure how well a pilot predicts work beyond the
observed prefix, we consider the 17 configurations where
neither checker finishes during its pilot and both
complete in the reference runs.
Table~\ref{tab:pilot-ratio-accuracy} reports the comparison.
The median relative difference is 6.62\%, and the mean
is 11.55\%.
The pilot estimate is within 10\% of the full-run value
on eleven configurations and within 20\% on thirteen.
The largest differences, about 42\% and 50\%, occur
on Barrier and Race-parametric.

\begin{table*}[t]
\centering
\tiny
\setlength{\tabcolsep}{10pt}
\renewcommand{\arraystretch}{1.05}
\begin{tabular}{lrrr}
\toprule
Configuration & $H_{\mathrm{pilot}}$ & $H_{\mathrm{full}}$ & Relative difference \\
\midrule
ApproxDS (N6, M4) & 10.401 & 9.755 & 6.62\% \\
Barrier (L2) & 3.386 & 2.388 & 41.81\% \\
DQ (L3) & 2.365 & 2.371 & 0.23\% \\
Mutex Musl (L1) & 2.124 & 2.126 & 0.08\% \\
Optimistic list (L2) & 5.189 & 5.736 & 9.53\% \\
QU (4,3,3) & 5.630 & 5.727 & 1.69\% \\
QU (3,3,3) & 5.407 & 5.326 & 1.51\% \\
Scull (R3, W2) & 6.993 & 7.665 & 8.76\% \\
Spinlock (N4, U6) & 3.169 & 3.549 & 10.71\% \\
STC (P5, 2/2/2/2, U16) & 4.817 & 4.848 & 0.63\% \\
Ticketlock (L1) & 2.607 & 3.382 & 22.92\% \\
Treiber dynamic (W4, RW1) & 2.084 & 2.789 & 25.29\% \\
Treiber static (L3) & 6.796 & 7.789 & 12.75\% \\
TTAS & 2.847 & 2.859 & 0.42\% \\
Scull (R2, W2) & 7.404 & 7.279 & 1.72\% \\
Sigma (N9) & 6.432 & 6.514 & 1.26\% \\
Race-parametric (N7) & 17.761 & 11.807 & 50.44\% \\
\midrule
\multicolumn{3}{l}{Median relative difference} & 6.62\% \\
\multicolumn{3}{l}{Mean relative difference} & 11.55\% \\
\bottomrule
\end{tabular}
\caption{RF/Maz cost per reported normal execution,
measured by pilots and complete checker runs.
$H_{\mathrm{pilot}}=1/\eta$ and
$H_{\mathrm{full}}=(T_R/N_R)/(T_M/N_M)$;
the relative difference is
$100|H_{\mathrm{pilot}}/H_{\mathrm{full}}-1|$.
Both checkers remain unfinished after their one-minute
pilots and complete in the reference runs on all 17 configurations.
Each row compares one recorded pilot pair with one full-run
pair, using actual observation times.}
\label{tab:pilot-ratio-accuracy}
\end{table*}

\pheading{How often do the samples support the choice?}
We use the test from Section~\ref{counting_algo} to assess
the evidence for our choices, at a nominal significance
level of 5\% per one-sided test.
With $b=20$ and $b=50$, the test is conclusive in
298 and 308 of 340 runs, respectively;
every conclusion agrees with the observed faster checker.
The remaining 42 and 32 runs include all seven and two
incorrect choices, respectively.
Ten runs in each group complete only one trial on
Opt-lock (N512), so the test cannot estimate the variance.
With $b=1$, all 18 incorrect choices likewise occur
among the 94 inconclusive runs.

\section{Related Work}
\label{sec:related}

In this paper, we study the size of the execution space under RF reduction
and its ratio to the space under Mazurkiewicz reduction.
\pheading{The counting class $\mathrm{SpanP}$.}
The class $\mathrm{SpanP}$~\cite{spanp},
equivalently $\#\!\cdot\!\mathrm{NP}$\cite{hemaspaandra1995satanic}, consists of
functions that count polynomial-size strings satisfying
an NP predicate.
RF counting belongs to this class: a nondeterministic
polynomial-time procedure can guess a maximal execution
and output its RF graph, counting distinct outputs. Toda and Watanabe~\cite{todaWatanabe1992} show that
every $\mathrm{SpanP}$ function can be computed in
polynomial time with access to a $\sharpP$ oracle.
Consequently, every $\sharpP$-hard function in
$\mathrm{SpanP}$ is $\mathrm{SpanP}$-complete under
polynomial-time Turing reductions.
Our hardness result therefore makes RF counting
$\mathrm{SpanP}$-complete under these reductions.
However, Mazurkiewicz counting is also
$\mathrm{SpanP}$-complete under the same reductions,
so this classification does not establish that RF
counting is harder.
Our conditional separation of RF counting from
$\sharpP$ in Section~\ref{sec:complexity} provides
this distinction.

\pheading{Model counting.}
RF counting can be formulated as projected model
counting~\cite{projected2015}: projection variables
describe an RF graph, while auxiliary variables encode
an execution witnessing its realizability.
The formula can be handled by exact projected counters
or hashing-based approximate counters such as
ApproxMC~\cite{approxmc2013,approxmc}, following
Stockmeyer's approach~\cite{stockmeyer1983}.

We pursue an execution-based approach for two reasons.
First, building a monolithic program encoding for estimation
would reintroduce a cost that execution-based checkers avoid
by evaluating local computations directly.
Second, hashing-based counters obtain their polynomial-time
guarantees relative to a SAT oracle.
Fixing the thread count $t$ does not bound the encoding's
variable count independently of program size or provide
a polynomial-time procedure for the resulting SAT queries.
These guarantees therefore do not recover our
polynomial-time-per-trial bound for fixed $t$.

Related algorithms count words and Mazurkiewicz traces
from explicit automata~\cite{meel2024fasterfprasnfa,ColnetMM26};
we estimate RF classes directly from bounded programs.

\pheading{Coarser execution equivalences.}
RF equivalence is itself part of a broader effort to reduce
the execution space beyond Mazurkiewicz equivalence.
Data-centric DPOR and optimal RF exploration exploit
read sources to merge otherwise distinct
classes~\cite{dcpor2018,rfsc2019}.
Value-based abstractions can merge executions further
by identifying reads that obtain the same value from
different writes.
In particular, Agarwal et al.~\cite{rvf2021} introduce
reads-value-from (RVF) equivalence, which preserves read
values and the causal ordering among reads, and develop
the RVF-SMC model checker. These approaches develop and explore coarser partitions;
our work asks how to quantify the reduction before
committing to a complete exploration.
Our concrete estimator and evaluation focus on RF
reduction relative to Mazurkiewicz reduction.
Extending checker selection to RVF would additionally
require estimating its partition and accounting for
the cost of its exploration procedure.

\section{Conclusion}

In this work, we studied how to quantify the advantage of RF equivalence
for individual programs without completing their exploration. We showed
that counting RF classes is $\sharpP$-hard and that even polynomial-time
unbiased estimation is likely impossible for an unbounded number of
threads. For a fixed number of threads, however, our algorithm \name{}
produces unbiased estimates of the RF and Mazurkiewicz class counts in
polynomial time per trial. The resulting ratio estimator is strongly
consistent.

Our experiments show that these estimates can help choose between RF and
Mazurkiewicz model checkers. Combining the estimated reduction ratio with
the exploration rates measured during short runs of the two checkers
reduces total verification time on our benchmarks, even after including
the time spent making the choice. Our comparison with canonical-hit
sampling shows that weighting improves both estimation accuracy and
checker selection.

\pheading{AI Use.}
We used OpenAI's GPT models to polish the paper,
assist with proofs, and support algorithmic discussions
and development.
We also used OpenAI's Codex and Anthropic's Claude Code
to assist with the implementation used in our experiments.

\section*{Data-Availability Statement}
The code and data underlying this study include
the implementation of \name{} in GenMC, benchmark
programs and configurations, experimental logs,
and scripts for running the experiments and generating
the reported tables.
We plan to make these materials publicly available.
\bibliographystyle{ACM-Reference-Format}
\bibliography{references}

@article{pldi26,
  author = {A. R. Balasubramanian and Mohammad Hossein {Khoshechin Jorshari} and Rupak Majumdar and Umang Mathur and Minjian Zhang},
  title = {State Space Estimation for {DPOR}-Based Model Checkers},
  journal = {Proceedings of the ACM on Programming Languages},
  volume = {10},
  number = {PLDI},
  articleno = {213},
  pages = {1102--1126},
  numpages = {25},
  year = {2026},
  month = jun,
  doi = {10.1145/3808291},
  url = {https://doi.org/10.1145/3808291}
}

@inproceedings{flanagan2005dpor,
author = {Flanagan, Cormac and Godefroid, Patrice},
title = {Dynamic partial-order reduction for model checking software},
year = {2005},
isbn = {158113830X},
publisher = {Association for Computing Machinery},
address = {New York, NY, USA},
doi = {10.1145/1040305.1040315},
booktitle = {Proceedings of the 32nd ACM SIGPLAN-SIGACT Symposium on Principles of Programming Languages},
pages = {110–121},
numpages = {12},
location = {Long Beach, California, USA},
series = {POPL '05}
}

@article{rfsc2019,
  author = {Parosh Aziz Abdulla and Mohamed Faouzi Atig and Bengt Jonsson and Magnus L{\aa}ng and Tuan Phong Ngo and Konstantinos Sagonas},
  title = {Optimal Stateless Model Checking for Reads-From Equivalence under Sequential Consistency},
  journal = {Proceedings of the ACM on Programming Languages},
  volume = {3},
  number = {OOPSLA},
  articleno = {150},
  numpages = {29},
  year = {2019},
  month = oct,
  doi = {10.1145/3360576},
  url = {https://user.it.uu.se/~bengt/Papers/Full/oopsla19.pdf}
}

@article{dcpor2018,
  author = {Marek Chalupa and Krishnendu Chatterjee and Andreas Pavlogiannis and Nishant Sinha and Kapil Vaidya},
  title = {Data-Centric Dynamic Partial Order Reduction},
  journal = {Proceedings of the ACM on Programming Languages},
  volume = {2},
  number = {POPL},
  articleno = {31},
  numpages = {31},
  year = {2018},
  month = jan,
  doi = {10.1145/3158119},
  url = {https://cs.au.dk/~pavlogiannis/publications/papers/popl18a.pdf}
}

@article{TruSt2022,
  author = {Michalis Kokologiannakis and Iason Marmanis and Vladimir Gladstein and Viktor Vafeiadis},
  title = {Truly Stateless, Optimal Dynamic Partial Order Reduction},
  journal = {Proceedings of the ACM on Programming Languages},
  volume = {6},
  number = {POPL},
  articleno = {49},
  numpages = {28},
  year = {2022},
  month = jan,
  doi = {10.1145/3498711},
  url = {https://plv.mpi-sws.org/genmc/popl2022-trust.pdf}
}

@article{knuth1975,
  author = {Donald E. Knuth},
  title = {Estimating the Efficiency of Backtrack Programs},
  journal = {Mathematics of Computation},
  volume = {29},
  number = {129},
  pages = {121--136},
  year = {1975},
  doi = {10.2307/2005469},
  url = {https://www.cs.brandeis.edu/~storer/JimPuzzles/MATCH/InstantInsanity/InstantInsanityKnuthArticle.pdf}
}

@article{vaisman2017,
  author = {Radislav Vaisman and Dirk P. Kroese},
  title = {Stochastic Enumeration Method for Counting Trees},
  journal = {Methodology and Computing in Applied Probability},
  volume = {19},
  number = {1},
  pages = {31--73},
  year = {2017},
  doi = {10.1007/s11009-015-9457-4},
  url = {https://people.smp.uq.edu.au/RadislavVaisman/papers/se-tree-counting.pdf}
}

@article{approxmc,
  author = {Mate Soos and Kuldeep S. Meel},
  title = {{BIRD}: Engineering an Efficient {CNF-XOR SAT} Solver and Its Applications to Approximate Model Counting},
  journal = {Proceedings of the AAAI Conference on Artificial Intelligence},
  volume = {33},
  number = {1},
  pages = {1592--1599},
  year = {2019},
  doi = {10.1609/aaai.v33i01.33011592},
  url = {https://www.cs.toronto.edu/~meel/Papers/aaai19-sm.pdf}
}

@inproceedings{approxmc2013,
  author = {Supratik Chakraborty and Kuldeep S. Meel and Moshe Y. Vardi},
  title = {A Scalable Approximate Model Counter},
  booktitle = {Principles and Practice of Constraint Programming},
  publisher = {Springer},
  address = {Berlin, Heidelberg},
  pages = {200--216},
  year = {2013},
  doi = {10.1007/978-3-642-40627-0_18},
  url = {https://www.cs.toronto.edu/~meel/Papers/CP2013.pdf}
}

@inproceedings{projected2015,
  author = {Rehan Abdul Aziz and Geoffrey Chu and Christian Muise and Peter Stuckey},
  title = {\#{\ensuremath{\exists}}{SAT}: Projected Model Counting},
  booktitle = {Theory and Applications of Satisfiability Testing---SAT 2015},
  publisher = {Springer},
  address = {Cham},
  series = {Lecture Notes in Computer Science},
  volume = {9340},
  pages = {121--137},
  year = {2015},
  doi = {10.1007/978-3-319-24318-4_10},
  url = {https://arxiv.org/abs/1507.07648}
}

@article{spanp,
  author = {Johannes K{\"o}bler and Uwe Sch{\"o}ning and Jacobo Tor{\'a}n},
  title = {On Counting and Approximation},
  journal = {Acta Informatica},
  volume = {26},
  number = {4},
  pages = {363--379},
  year = {1989},
  doi = {10.1007/BF00276023},
  url = {https://link.springer.com/article/10.1007/BF00276023}
}

@article{gibbons1997,
  author = {Phillip B. Gibbons and Ephraim Korach},
  title = {Testing Shared Memories},
  journal = {SIAM Journal on Computing},
  volume = {26},
  number = {4},
  pages = {1208--1244},
  year = {1997},
  doi = {10.1137/S0097539794279614},
  url = {https://epubs.siam.org/doi/10.1137/S0097539794279614}
}

@article{todaWatanabe1992,
  author  = {Seinosuke Toda and Osamu Watanabe},
  title   = {Polynomial-Time 1-Turing Reductions
             from {\#PH} to {\#P}},
  journal = {Theoretical Computer Science},
  volume  = {100},
  number  = {1},
  pages   = {205--221},
  year    = {1992},
  doi     = {10.1016/0304-3975(92)90369-Q}
}

@article{fenner1994,
  author  = {Stephen A. Fenner and Lance J. Fortnow
             and Stuart A. Kurtz},
  title   = {Gap-Definable Counting Classes},
  journal = {Journal of Computer and System Sciences},
  volume  = {48},
  number  = {1},
  pages   = {116--148},
  year    = {1994},
  doi     = {10.1016/S0022-0000(05)80024-8}
}

@inproceedings{stockmeyer1983,
  author    = {Larry Stockmeyer},
  title     = {The Complexity of Approximate Counting},
  booktitle = {Proceedings of the Fifteenth Annual ACM
               Symposium on Theory of Computing},
  publisher = {Association for Computing Machinery},
  pages     = {118--126},
  year      = {1983},
  doi       = {10.1145/800061.808740}
}

@article{fenner2003awpp,
  author  = {Stephen A. Fenner},
  title   = {{PP}-Lowness and a Simple Definition of {AWPP}},
  journal = {Theory of Computing Systems},
  volume  = {36},
  number  = {2},
  pages   = {199--212},
  year    = {2003},
  doi     = {10.1007/s00224-002-1089-8}
}

@article{deCampos2020weighted,
  author  = {de Campos, Cassio P. and
             Stamoulis, Georgios and
             Weyland, Dennis},
  title   = {A structured view on weighted counting with
             relations to counting, quantum computation
             and applications},
  journal = {Information and Computation},
  volume  = {275},
  pages   = {104627},
  year    = {2020},
  doi     = {10.1016/j.ic.2020.104627}
}

@inproceedings{genmc2021,
  author = {Michalis Kokologiannakis and Viktor Vafeiadis},
  title = {{GenMC}: A Model Checker for Weak Memory Models},
  booktitle = {Computer Aided Verification},
  series = {Lecture Notes in Computer Science},
  volume = {12759},
  pages = {427--440},
  publisher = {Springer},
  year = {2021},
  doi = {10.1007/978-3-030-81685-8_20}
}

@misc{genmc-benchmarks,
  author = {{GenMC Developers}},
  title = {{GenMC} v0.7 Regression Benchmarks},
  year = {2022},
  url = {https://github.com/MPI-SWS/genmc/tree/v0.7/tests/correct/data-structures},
  note = {Release commit 0f94ce362c9d91e269f9dcd66791792b8bb8eb2c; accessed 22 September 2026}
}

@book{book:herlihy-shavit,
  author = {Maurice Herlihy and Nir Shavit and Victor Luchangco and Michael Spear},
  title = {The Art of Multiprocessor Programming},
  edition = {2},
  publisher = {Morgan Kaufmann},
  year = {2020},
  doi = {10.1016/C2011-0-06993-4}
}

@misc{nidhugg-benchmarks,
  author = {{Nidhugg Developers}},
  title = {{Nidhugg} Benchmark Suite},
  url = {https://github.com/nidhugg/nidhugg/tree/master/benchmarks},
  note = {approxds\_append.c and the RFSC/DC-DPOR Parker benchmark; accessed 22 September 2026}
}

@article{farzan2024coarser,
  author    = {Azadeh Farzan and Umang Mathur},
  title     = {Coarser Equivalences for Causal Concurrency},
  journal   = {Proceedings of the ACM on Programming Languages},
  volume    = {8},
  number    = {POPL},
  articleno = {31},
  numpages  = {31},
  year      = {2024},
  doi       = {10.1145/3632873}
}

@article{papadimitriou1979,
  author  = {Christos H. Papadimitriou},
  title   = {The Serializability of Concurrent Database Updates},
  journal = {Journal of the ACM},
  volume  = {26},
  number  = {4},
  pages   = {631--653},
  year    = {1979},
  doi     = {10.1145/322154.322158}
}

@article{fortnowGrochow2011,
  author  = {Lance Fortnow and Joshua A. Grochow},
  title   = {Complexity Classes of Equivalence Problems Revisited},
  journal = {Information and Computation},
  volume  = {209},
  number  = {4},
  pages   = {748--763},
  year    = {2011},
  doi     = {10.1016/j.ic.2011.01.006}
}

@InProceedings{observers2018,
author="Aronis, Stavros
and Jonsson, Bengt
and L{\aa}ng, Magnus
and Sagonas, Konstantinos",
editor="Beyer, Dirk
and Huisman, Marieke",
title="Optimal Dynamic Partial Order Reduction with Observers",
booktitle="Tools and Algorithms for the Construction and Analysis of Systems",
year="2018",
publisher="Springer International Publishing",
address="Cham",
pages="229--248",
isbn="978-3-319-89963-3"
}

@article{vcpor2019,
author = {Chatterjee, Krishnendu and Pavlogiannis, Andreas and Toman, Viktor},
title = {Value-centric dynamic partial order reduction},
year = {2019},
issue_date = {October 2019},
publisher = {Association for Computing Machinery},
address = {New York, NY, USA},
volume = {3},
number = {OOPSLA},
doi = {10.1145/3360550},
journal = {Proc. ACM Program. Lang.},
month = oct,
articleno = {124},
numpages = {29}
}

@inproceedings{rvf2021,
author = {Agarwal, Pratyush and Chatterjee, Krishnendu and Pathak, Shreya and Pavlogiannis, Andreas and Toman, Viktor},
title = {Stateless Model Checking Under a Reads-Value-From Equivalence},
year = {2021},
isbn = {978-3-030-81684-1},
publisher = {Springer-Verlag},
address = {Berlin, Heidelberg},
url = {https://doi-org.libproxy1.nus.edu.sg/10.1007/978-3-030-81685-8_16},
doi = {10.1007/978-3-030-81685-8_16},
booktitle = {Computer Aided Verification: 33rd International Conference, CAV 2021, Virtual Event, July 20–23, 2021, Proceedings, Part I},
pages = {341–366},
numpages = {26}
}

@article{ColnetMM26,
  author = {de Colnet, Alexis and Meel, Kuldeep S.
            and Mathur, Umang},
  title = {Counting and Sampling Traces in Regular Languages},
  journal = {Proceedings of the ACM on Programming Languages},
  volume = {10},
  number = {POPL},
  articleno = {81},
  pages = {2352--2379},
  numpages = {28},
  year = {2026},
  month = jan,
  doi = {10.1145/3776723},
  url = {https://doi.org/10.1145/3776723}
}

@InProceedings{awamoche,
author="Kokologiannakis, Michalis
and Marmanis, Iason
and Vafeiadis, Viktor",
editor="Enea, Constantin
and Lal, Akash",
title="Unblocking Dynamic Partial Order Reduction",
booktitle="Computer Aided Verification",
year="2023",
publisher="Springer Nature Switzerland",
address="Cham",
pages="230--250",
isbn="978-3-031-37706-8"
}

@article{hemaspaandra1995satanic,
  author  = {Lane A. Hemaspaandra and Heribert Vollmer},
  title   = {The Satanic Notations: Counting Classes beyond
             {\#P} and Other Definitional Adventures},
  journal = {ACM SIGACT News},
  volume  = {26},
  number  = {1},
  pages   = {2--13},
  year    = {1995},
  doi     = {10.1145/203610.203611}
}

@article{meel2024fasterfprasnfa,
  author  = {Kuldeep S. Meel and Sourav Chakraborty
             and Umang Mathur},
  title   = {A Faster {FPRAS} for {\#NFA}},
  journal = {Proceedings of the ACM on Management of Data},
  volume  = {2},
  number  = {2},
  articleno = {112},
  numpages = {22},
  year    = {2024},
  doi     = {10.1145/3651613}
}

\clearpage

\appendix
\section{Proof of Theorem~\ref{thm:rf-sharpp-hard}}
\label{app:rf-sharpp-hard}

\paragraph{Proof idea.}
We reuse the reduction establishing
$\sharpP$-hardness of Mazurkiewicz counting in
Balasubramanian et al.~\cite{pldi26}.
The constructed programs satisfy
$C_{\RF}(P)=C_{\Maz}(P)$, so the same reduction establishes
$\sharpP$-hardness of RF counting.

\section{Proof of Proposition~\ref{thm:ratio-hardness}}
\label{app:ratio-hardness}

\paragraph{Proof idea.}
Given a Boolean formula $\varphi$ over $n$ variables,
construct a program with shared locations
$x_1,\ldots,x_n,z$, initially zero.
For each $i$, a separate thread writes $x_i:=1$.
A main thread reads each $x_i$ into a local bit $b_i$,
evaluates $\varphi(b_1,\ldots,b_n)$, and writes $z:=1$
if the formula holds.
One additional thread writes $z:=2$.
No thread reads $z$.

The reads can realize every Boolean assignment, and each
assignment determines exactly one RF class.
An unsatisfying assignment contributes one Mazurkiewicz
class.
A satisfying assignment contributes two,
distinguished by the order of the two writes to $z$.
Hence,
\[
  C_{\RF}(P_\varphi)=2^n,
  \qquad
  C_{\Maz}(P_\varphi)=2^n+\#\mathrm{SAT}(\varphi).
\]
The exact ratio therefore determines the model count:
\[
  \#\mathrm{SAT}(\varphi)
  =
  2^n\left(\frac{1}{\rho(P_\varphi)}-1\right).
\]
This establishes $\sharpP$-hardness with a single oracle
query.

\section{Proof of Theorem~\ref{thm:rf-not-sharpp}}
\label{app:sat-gap}

\begin{proof}
We prove that $C_{\RF}\in\sharpP$ implies
$\mathrm{NP}\subseteq\mathrm{SPP}$.
The key is a polynomial-time construction that maps a Boolean
formula $\varphi$ to two loop-free deterministic SC read/write
programs $P_1$ and $P_2$ satisfying
\begin{equation}
 C_{\RF}(P_2)-C_{\RF}(P_1)=
 \begin{cases}
  1,&\text{if $\varphi$ is satisfiable},\\
  0,&\text{otherwise}.
 \end{cases}
 \label{eq:sat-count-difference}
\end{equation}
All counts range over maximal executions.

Our construction combines Papadimitriou's SAT reduction for
polygraph acyclicity~\cite[Section~3, Theorem~1]{papadimitriou1979}
with the serialization gadget used by Gibbons and Korach for
SC verification with a fixed read-mapping~\cite[Section~4.1,
Theorem~4.1]{gibbons1997}.
We implement the prescribed read sources using guarded reads
and add an auxiliary read to turn realizability into a difference
of class counts.

\pheading{The source formula.}
A clause is \emph{mixed} if it contains both positive and negative
literals. A formula is \emph{noncircular} if at most one occurrence
of each variable belongs to a mixed clause.
By Papadimitriou's normalization~\cite[Lemma~3]{papadimitriou1979},
we can transform $\varphi$ in polynomial time into an
equisatisfiable noncircular CNF formula $F$ with two or three
literals per clause.
We use the variable and clause gadgets of that reduction below.

\pheading{Blocks and prescribed read sources.}
The reduction encodes Boolean choices and clause constraints
through the relative order of blocks.
To implement these constraints using ordinary SC reads and
writes, we ensure that block bodies do not interleave in any
execution where all success checks pass.
Each block $T$ is implemented by a separate thread with a distinct
nonzero identifier $\operatorname{id}(T)$.
All shared locations are initially zero.
Let $A$ be a common marker and let $\mathit{finish}_T$ be a fresh
shared flag written only by $T$.
The block executes
\begin{lstlisting}[style=paperprogram]
(*@$\mathtt{ok}_T$@*) (*@$\gets$@*) true;
write(A, id((*@$T$@*)));
body((*@$T$@*));
a (*@$\gets$@*) read(A);
if (*@$\mathtt{ok}_T$@*) (*@$\land$@*) a = id((*@$T$@*)) then
    write((*@$\mathtt{finish}_T$@*), 1);
\end{lstlisting}
The local flag $\mathit{ok}_T$ records whether all prescribed reads
in the body succeeded.
A failed check sets it to false, but does not stop the block or
skip subsequent body operations.

If every block writes its finish flag, every marker read must
read its own block's marker write.
The intervals from these writes to their corresponding reads
are pairwise disjoint: if two overlapped, the later marker write
would prevent the earlier block's read from reading its own write.
Successful blocks therefore have a strict total order,
denoted by $<$.

We use two operations to construct the block bodies.
The operation $\operatorname{RequireRF}(T,U)$ introduces a fresh
location $e$: block $T$ writes $\operatorname{id}(T)$ to $e$,
and block $U$ reads $e$ and checks that it obtains this value.
An accompanying operation
$\operatorname{AddCompetingWrite}(T,U,W)$ adds
$\operatorname{write}(e,\operatorname{id}(W))$
to a third block $W$.
Each call to $\operatorname{RequireRF}$ uses a fresh location;
the accompanying operation refers to that same location.
Within each block, fix a deterministic order for all added
operations.

Without the competing write, a successful prescribed read requires
$T<U$.
With the competing write, it requires precisely
\begin{equation}
 T<U\quad\text{and}\quad(W<T\ \lor\ U<W).
 \label{eq:sat-block-order}
\end{equation}
The second condition prevents $W$ from overwriting the designated
source between $T$ and $U$.
Conversely, executing the blocks serially in any total order
satisfying these constraints makes every prescribed read succeed.

\pheading{Variable and literal gadgets.}
For each variable $q$ of $F$, create three blocks $a_q,b_q,c_q$
and add
\[
 \operatorname{RequireRF}(a_q,b_q),\qquad
 \operatorname{AddCompetingWrite}(a_q,b_q,c_q).
\]
Their two possible successful orders encode the truth value of $q$:
\[
 q=1:\ c_q<a_q<b_q,
 \qquad
 q=0:\ a_q<b_q<c_q.
\]

For each literal occurrence $\ell$, create fresh blocks
$y_\ell,z_\ell$.
For a positive occurrence $\ell=q$, add
\[
 \operatorname{RequireRF}(c_q,y_\ell),\qquad
 \operatorname{RequireRF}(b_q,z_\ell),\qquad
 \operatorname{AddCompetingWrite}(b_q,z_\ell,y_\ell).
\]
For a negative occurrence $\ell=\neg q$, add
\[
 \operatorname{RequireRF}(z_\ell,c_q),\qquad
 \operatorname{RequireRF}(y_\ell,a_q),\qquad
 \operatorname{AddCompetingWrite}(y_\ell,a_q,z_\ell).
\]
These constraints force $z_\ell<y_\ell$ whenever $\ell$
evaluates to false.
If $\ell=q$ is false, then $q=0$ and hence
$b_q<c_q<y_\ell$.
This rules out $y_\ell<b_q$ in~\eqref{eq:sat-block-order},
forcing $z_\ell<y_\ell$.
If $\ell=\neg q$ is false, then $q=1$ and hence
$z_\ell<c_q<a_q$.
This rules out $a_q<z_\ell$, again forcing $z_\ell<y_\ell$.

\pheading{Clause gadgets and soundness.}
For a clause with occurrences $\ell_1,\ldots,\ell_k$,
where $k\in\{2,3\}$, add
\[
 \operatorname{RequireRF}(y_{\ell_i},z_{\ell_{i+1}})
 \quad(1\le i\le k),\qquad \ell_{k+1}=\ell_1.
\]
If every literal in the clause were false, the successful block
order would contain the cycle
\[
 z_{\ell_1}<y_{\ell_1}<z_{\ell_2}<y_{\ell_2}
 <\cdots<z_{\ell_k}<y_{\ell_k}<z_{\ell_1},
\]
which is impossible.
Thus, any execution in which every block writes its finish flag
determines a satisfying assignment of $F$.

\pheading{Completeness.}
Conversely, fix a satisfying assignment of $F$.
For each false literal choose $z_\ell<y_\ell$ in its ordering
disjunction.
For a true positive literal choose $y_\ell<b_q$;
for a true negative literal choose $a_q<z_\ell$.
Choose the variable orders according to the assignment.

These are exactly the variable- and clause-gadget edge choices
in Papadimitriou's SAT-to-polygraph construction.
Its completeness argument shows that, for a noncircular formula,
the resulting directed graph is acyclic~
\cite[proof of Theorem~1]{papadimitriou1979}.
A topological order therefore gives a total block order satisfying
all constraints in~\eqref{eq:sat-block-order}.
Executing complete blocks in this order makes every marker and
prescribed body read succeed, so every finish flag is written.

\pheading{The two programs.}
Program $P_1$ consists of the block threads and one checker thread.
The checker reads every $\mathit{finish}_T$ once in a fixed order,
without short-circuiting.
If all these reads return $1$, it performs one additional read
of a fresh location $\mathit{aux}$, initially $0$.
It discards that read's value and terminates.
Call an execution \emph{successful} when this auxiliary read occurs.

Program $P_2$ is identical, except for one additional thread
whose only operation is $\operatorname{write}(\mathit{aux},1)$.
All threads are loop-free and nonblocking, so every maximal
execution completes every thread.

By soundness, a successful execution yields a satisfying
assignment of $F$.
By completeness, a satisfying assignment gives a successful
execution by running all blocks in the constructed order and then
running the checker.
Consequently, a successful execution exists exactly when
$\varphi$ is satisfiable.

\pheading{One successful RF class.}
Existence alone is not enough for~\eqref{eq:sat-count-difference}:
we also need all successful executions of $P_1$ to induce the
same RF graph.
In every such execution, all body events and finish writes occur.
Each body read has its prescribed source; each marker read reads
its own block's marker write; each checker read of a finish flag
reads its unique noninitialization write; and the auxiliary read
reads initialization.
Thus, the event set, event labels, and all read sources are fixed.
Different satisfying assignments and block orders only change
information discarded by RF projection.
Hence the number of successful RF classes of $P_1$ is exactly
one if $\varphi$ is satisfiable and zero otherwise.

\pheading{Cancellation of the other classes.}
Map each RF graph of $P_2$ to one of $P_1$ by deleting the extra
auxiliary write.
If the auxiliary read reads from that write, redirect it to
initialization and change its returned value to $0$.
Deleting the write from an SC execution witnesses realizability
of the resulting graph: no other operation accesses this fresh
location, and the read's value has no effect on subsequent behavior.

Every unsuccessful RF graph of $P_1$ has exactly one preimage:
there is no auxiliary read, so the additional write introduces
no read-source choice.
This preimage is realizable by scheduling the write after all
other events.

Every successful RF graph has exactly two preimages, according
to whether the auxiliary read reads initialization or the added
write.
The former is realized by placing the write last; the latter
by placing it immediately before the auxiliary read.
These are the only possible sources, and all other graph
components are unchanged.
Therefore,
\[
 C_{\RF}(P_2)-C_{\RF}(P_1)
 =
 \bigl|\{\text{successful RF classes of }P_1\}\bigr|,
\]
which proves~\eqref{eq:sat-count-difference}.

\pheading{Size of the construction.}
The number of blocks, locations, and instructions is polynomial
in $|\varphi|$.
Block identifiers require only logarithmically many bits,
and local computation consists of comparisons and Boolean
operations.
Thus, both programs, including their execution-length bounds,
are constructible in polynomial time.

\pheading{Consequence for $\sharpP$ membership.}
Suppose $C_{\RF}\in\sharpP$.
Since $P_1(\varphi)$ and $P_2(\varphi)$ are constructible
in polynomial time, both functions
\[
 \varphi\mapsto C_{\RF}(P_1(\varphi)),
 \qquad
 \varphi\mapsto C_{\RF}(P_2(\varphi))
\]
belong to $\sharpP$.
Their difference therefore belongs to $\mathrm{GapP}$.
By~\eqref{eq:sat-count-difference}, this difference is the
characteristic function of SAT: it equals $1$ on satisfiable
formulas and $0$ on unsatisfiable formulas.
The characterization of $\mathrm{SPP}$ by $0$--$1$-valued
$\mathrm{GapP}$ functions~\cite{fenner1994} then gives
$\mathrm{SAT}\in\mathrm{SPP}$.
Since SAT is NP-complete and $\mathrm{SPP}$ is closed under
polynomial-time many-one reductions, we obtain
$\mathrm{NP}\subseteq\mathrm{SPP}$.
Thus, RF counting does not belong to $\sharpP$ unless
$\mathrm{NP}\subseteq\mathrm{SPP}$.
\end{proof}

\section{Proof of Theorem~\ref{thm:rf-unbiased-barrier}}
\label{app:rf-unbiased-barrier}

\begin{proof}
Suppose, for contradiction, that an estimator $A$ satisfying
the conditions of the theorem exists.
We show that $3$-SAT is low for $\mathrm{PP}$, contradicting
the assumption $\mathrm{PP}^{\mathrm{NP}}\neq\mathrm{PP}$.
The weighted-counting argument below is an explicit
specialization of the arguments underlying
Theorems~4, 8, and~17 of de Campos et al.~
\cite{deCampos2020weighted},
combined with Fenner's PP-lowness criterion~
\cite[Theorem~1.1]{fenner2003awpp}.

Recall that $\mathrm{GapP}$ consists of differences of two
$\sharpP$ functions.
Fenner's criterion states that a language $L$ is low for
$\mathrm{PP}$ if there exist a function $h\in\mathrm{GapP}$
and a polynomial $q$ such that, for every input $x$,
\[
  \frac{h(x)}{2^{q(|x|)}}
  \in
  \begin{cases}
    [0,1/3], & x\notin L,\\[2pt]
    [2/3,1], & x\in L.
  \end{cases}
\]
Here, lowness means $\mathrm{PP}^{L}=\mathrm{PP}$.
We construct such a function for $3$-SAT using the
assumed unbiased estimator.

All constructions below are specified for valid $3$-CNF
encodings and extended by zero to malformed inputs.
These extensions can be implemented by polynomial-time
validity checks and preserve the counting-class
membership claims below.

Let $\varphi$ be a $3$-CNF formula.
By the construction establishing
\eqref{eq:sat-count-difference}, there are polynomial-time
constructible programs $P_1(\varphi)$ and $P_2(\varphi)$
such that
\[
  \CRF(P_2(\varphi))-\CRF(P_1(\varphi))
  =\chi_{\mathrm{SAT}}(\varphi),
\]
where $\chi_{\mathrm{SAT}}(\varphi)$ is $1$ if $\varphi$
is satisfiable and $0$ otherwise.

Both programs have size polynomial in $|\varphi|$.
The worst-case polynomial running-time bound on $A$
therefore gives a polynomial $m=m(|\varphi|)$ bounding
the number of random bits used by either invocation.
Padding with unused random bits, we represent both runs
using a common uniformly distributed string
$r\in\{0,1\}^{m}$ and define
\[
  D_\varphi(r)
  =
  A(P_2(\varphi);r)-A(P_1(\varphi);r).
\]
Each invocation has the correct marginal distribution.
Thus, unbiasedness and linearity of expectation give
\[
  \mathbb E_r[D_\varphi(r)]
  =\chi_{\mathrm{SAT}}(\varphi).
\]

To obtain an integer-valued function, define
\[
  b_\varphi(r)
  =
  \left\lfloor 8D_\varphi(r)+\frac12\right\rfloor.
\]
Because $A$ outputs explicit binary rational numbers
within polynomial time, $b_\varphi(r)$ is computable
in polynomial time and has polynomial bit length.
Moreover,
\[
  \left|b_\varphi(r)-8D_\varphi(r)\right|
  \leq \frac12.
\]
Taking expectations yields
\[
  \left|
    \mathbb E_r[b_\varphi(r)]
    -8\chi_{\mathrm{SAT}}(\varphi)
  \right|
  \leq \frac12.
\]

Now define
\[
  g(\varphi)
  =
  \sum_{r\in\{0,1\}^{m}} b_\varphi(r).
\]
We claim that $g\in\mathrm{GapP}$.
Write
\[
  b_\varphi^+(r)=\max\{b_\varphi(r),0\},
  \qquad
  b_\varphi^-(r)=\max\{-b_\varphi(r),0\}.
\]
Choose a polynomial $s$ such that
$b_\varphi^\pm(r)<2^{s(|\varphi|)}$ for every $r$.
Each sum $\sum_r b_\varphi^\pm(r)$ belongs to $\sharpP$:
a machine guesses $r$ and an $s(|\varphi|)$-bit
nonnegative integer $j$, and accepts exactly when
$j<b_\varphi^\pm(r)$.
Their difference is $g$, proving the claim.

Since $r$ is uniform,
\[
  \frac{g(\varphi)}{2^m}
  =\mathbb E_r[b_\varphi(r)].
\]
Consequently,
\[
  \frac{g(\varphi)}{2^m}
  \in
  \begin{cases}
    [-1/2,1/2],
      & \text{if $\varphi$ is unsatisfiable},\\[2pt]
    [15/2,17/2],
      & \text{if $\varphi$ is satisfiable}.
  \end{cases}
\]

To place these intervals within those required by
Fenner's criterion, define
\[
  h(\varphi)=g(\varphi)+4\cdot 2^m.
\]
The added term $2^{m+2}$ is a $\sharpP$ function:
a machine guesses $m+2$ bits and accepts on every path.
Thus $h\in\mathrm{GapP}$, and
\[
  \frac{h(\varphi)}{2^{m+4}}
  =\frac14+\frac{\mathbb E_r[b_\varphi(r)]}{16}
  \in
  \begin{cases}
    [7/32,9/32],
      & \text{if $\varphi$ is unsatisfiable},\\[2pt]
    [23/32,25/32],
      & \text{if $\varphi$ is satisfiable}.
  \end{cases}
\]
These intervals lie in $[0,1/3]$ and $[2/3,1]$,
respectively.
Taking $q(n)=m(n)+4$, Fenner's criterion therefore gives
\[
  \mathrm{PP}^{3\text{-}\mathrm{SAT}}=\mathrm{PP}.
\]
Since $3$-SAT is NP-complete under polynomial-time
many-one reductions, it follows that
$\mathrm{PP}^{\mathrm{NP}}=\mathrm{PP}$,
contradicting the assumption of the theorem.
\end{proof}

\label{app:rf-not-sharpp}

\section{Proof of Corollary~\ref{cor:ratio-unbiased-barrier}}
\label{app:ratio-unbiased-barrier}

\begin{proof}
Suppose such an estimator $Z_\rho(P)$ exists.
Independently generate a polynomial-time unbiased
Mazurkiewicz estimate $Z_{\Maz}(P)$.
Their product satisfies
\[
  \mathbb{E}[Z_\rho(P)Z_{\Maz}(P)]
  =
  \rho(P)\,C_{\Maz}(P)
  =
  C_{\RF}(P),
\]
contradicting Theorem~\ref{thm:rf-unbiased-barrier}.
\end{proof}

\section{Proof of Proposition~\ref{prop:canonical-completion}}
\label{app:canonical-completion}

\begin{proof}[Proof sketch]
The SC consistency procedure of
Abdulla et al.~\cite{rfsc2019} searches for an execution
realizing $H$, whose write orders determine a compatible
coherence order.
For every fixed $t$, this search takes polynomial time.
We fix a deterministic search order that depends only
on $H$.
If the search succeeds, $\kappa(H)$ is the full graph
obtained by adding the returned coherence order to $H$;
otherwise, $\kappa(H)=\bot$.
Since all search choices depend only on $H$, the
procedure always selects the same full graph for
the same RF graph.
\end{proof}

\section{Proof of Proposition~\ref{prop:rf-fixed-t-sharpp}}
\label{app:rf-fixed-t-sharpp}

\begin{proof}[Proof sketch]
A nondeterministic machine guesses an RF graph $H$
using a unique encoding and computes $G=\kappa(H)$.
It rejects if $G=\bot$.
Otherwise, it accepts if and only if $G$ is a valid
maximal execution of $P$ under SC.
For fixed $t$, these checks take polynomial time,
and the accepting computations correspond one-to-one
with RF classes.
Thus, RF counting belongs to $\sharpP$.

An unbiased estimator follows by randomly simulating
this machine and returning the inverse sampling
probability if the computation accepts, or zero
otherwise.
Each accepting computation contributes one in
expectation, and each trial takes polynomial time.
\end{proof}
\section{Proof of Proposition~\ref{prop:paired-estimation}}
\label{app:paired-estimation}

\begin{proof}
Recall that TruSt constructs execution graphs incrementally
through forward extensions and backward revisits,
while exploring each maximal Mazurkiewicz class
exactly once~\cite{TruSt2022}.
With a fixed scheduler, its exploration induces the
transition system $\mathcal D(P)$ described by
Balasubramanian et al.~\cite{pldi26}.
The states include the execution graph and the exploration
metadata required by TruSt.
Crucially, the exploration is branch-independent:
the successors of a state do not depend on which
other branches have previously been explored.
Thus, $\mathcal D(P)$ defines a fixed rooted exploration
tree that can be sampled without constructing it in full.

Let $\mathcal L$ denote its leaves and
$\mathcal L_{\max}\subseteq\mathcal L$ those representing
maximal executions of $P$.
Optimality gives a bijection between $\mathcal L_{\max}$
and the maximal Mazurkiewicz classes.
Other exploration leaves contribute zero, including
blocked leaves that do not represent maximal executions.
We refer to~\cite{pldi26} for the transition rules and
the polynomial bounds on path length and successor
generation.

Recall that Knuth's method gives an unbiased estimate of the total
leaf weight of a rooted tree~\cite{knuth1975}.
It samples a root-to-leaf path using known positive
child-selection probabilities and returns the sampled
leaf's weight divided by the product of these probabilities.
At a node with $d$ children, each choice can be implemented
using $\lceil\log_2 d\rceil$ fair random bits:
interpret the bits as a nonnegative integer $J$ and select
child $1+(J\bmod d)$.

TeStor uses stochastic enumeration, an extension of
Knuth's method that samples multiple paths while
preserving unbiasedness~\cite{pldi26}.
This extension does not affect the argument below:
both methods are unbiased for any fixed assignment
of leaf weights.

Our algorithm changes only these weights.
For Mazurkiewicz counting, each leaf representing
a maximal execution has weight one, and all other
leaves have weight zero.
For RF counting, we additionally set the weights
of noncanonical leaves to zero.
The canonicality checker selects exactly one full
graph for each RF class, and optimality ensures
that this graph appears at exactly one leaf.
The modified tree therefore has total leaf weight
$C_{\RF}(P)$.
Applying the same sampling procedure to the original
and modified weights yields
\[
  \mathbb E[X]=C_{\Maz}(P),
  \qquad
  \mathbb E[Y]=C_{\RF}(P).
\]
The canonicality procedure selects one valid full graph
for each realizable RF projection, independently of
the sampled graph and previous trials.
Since every such full graph occurs exactly once in
$\mathcal L_{\max}$, exactly one leaf contributes
for each maximal RF class.
Consequently,
\[
  \sum_{\ell\in\mathcal L}c_{\RF}(\ell)=C_{\RF}(P).
\]

The procedure $\pairedestimation$ applies these
two cost functions to the same sampled leaf, returning
\[
  X=\frac{c_{\Maz}(\ell)}{q(\ell)},
  \qquad
  Y=\frac{c_{\RF}(\ell)}{q(\ell)}.
\]
Thus,
\[
  \mathbb E[X]=C_{\Maz}(P),
  \qquad
  \mathbb E[Y]=C_{\RF}(P).
\]
Equivalently, the RF estimate retains the original
Mazurkiewicz weight on canonical leaves and replaces
it by zero on all other leaves.
The stochastic-enumeration extension of TeStor is
likewise unbiased for any fixed leaf-cost function~
\cite{pldi26}.
Applying it to $c_{\Maz}$ and $c_{\RF}$ therefore
preserves both conclusions.

Finally, let $(X_i,Y_i)$ be independent repetitions
of the paired trial.
The strong law of large numbers gives
\[
  \frac1n\sum_{i=1}^{n}X_i
    \xrightarrow{\mathrm{a.s.}} C_{\Maz}(P),
  \qquad
  \frac1n\sum_{i=1}^{n}Y_i
    \xrightarrow{\mathrm{a.s.}} C_{\RF}(P).
\]
Since $C_{\Maz}(P)>0$, the denominator is almost surely
positive for all sufficiently large $n$, and
\[
  \frac{\sum_{i=1}^{n}Y_i}{\sum_{i=1}^{n}X_i}
  \xrightarrow{\mathrm{a.s.}}
  \frac{C_{\RF}(P)}{C_{\Maz}(P)}
  =\rho(P).
\]
Any fixed convention used when the denominator is zero
does not affect this limit.
\end{proof}

\section{Inapproximability of the Reduction Ratio}
\label{app:ratio-no-fpras}

\begin{proposition}
\label{prop:ratio-no-fpras}
Unless $\mathrm{NP}=\mathrm{RP}$, there is no fully polynomial
randomized approximation scheme for
\[
  \rho(P)=\frac{C_{\mathrm{RF}}(P)}{C_{\mathrm{Maz}}(P)},
\]
even for loop-free deterministic SC read/write programs
with two threads and Boolean shared variables.
Both counts range over classes of maximal executions.
\end{proposition}

\begin{proof}
We adapt the assignment-and-amplification construction used
for Mazurkiewicz counting in~\cite{pldi26}.

Let $\varphi$ be a CNF formula over $n$ Boolean variables,
and let $s$ be its number of satisfying assignments.
Set
\[
  N=2^n,\qquad m=n+4,\qquad K=\binom{2m}{m}.
\]
Construct a program $P_\varphi$ with shared variables
$x_1,\ldots,x_n,z$, all initially zero, and the following
two threads.
The variables $a_1,\ldots,a_n$ are local to $T_0$.
\par\medskip
\noindent
\begin{minipage}[t]{0.48\linewidth}
\centering\textbf{Thread $T_0$}\par
\begin{lstlisting}[style=paperprogram]
(*@$\mathtt{a}_1 \gets \mathtt{x}_1$@*);
(*@$\vdots$@*)
(*@$\mathtt{a}_n \gets \mathtt{x}_n$@*);
if (*@$\varphi(\mathtt{a}_1,\ldots,\mathtt{a}_n)$@*) then
    (*@$\underbrace{\mathtt{z}\gets\mathtt{1};\ \cdots;\ \mathtt{z}\gets\mathtt{1};}_{m\text{ stores}}$@*)
\end{lstlisting}
\end{minipage}\hfill
\begin{minipage}[t]{0.48\linewidth}
\centering\textbf{Thread $T_1$}\par
\begin{lstlisting}[style=paperprogram]
(*@$\mathtt{x}_1 \gets \mathtt{1}$@*);
(*@$\vdots$@*)
(*@$\mathtt{x}_n \gets \mathtt{1}$@*);

(*@$\underbrace{\mathtt{z}\gets\mathtt{0};\ \cdots;\ \mathtt{z}\gets\mathtt{0};}_{m\text{ stores}}$@*)
\end{lstlisting}
\end{minipage}
\par\medskip

\emph{Encoding assignments.}
For each $i$, let $r_i$ denote the read of $x_i$ in $T_0$
and $w_i$ its unique noninitial write in $T_1$.
The value of $a_i$ is one exactly when $w_i$ precedes $r_i$.
Every assignment $a\in\{0,1\}^n$ is realizable:
process the indices in increasing order, scheduling
$w_i;r_i$ when $a_i=1$ and $r_i;w_i$ when $a_i=0$.
This respects both threads' program orders.
After these pairs have completed, both threads have finished
their accesses to $x_1,\ldots,x_n$.

For a fixed assignment, the reads-from sources of all reads
are fixed, as is the set of executed events.
There are no reads of $z$.
Consequently, all executions producing that assignment belong
to the same RF class.
Different assignments have different reads-from relations,
so
\[
  C_{\mathrm{RF}}(P_\varphi)=N.
\]

\emph{Counting Mazurkiewicz classes.}

If an assignment does not satisfy $\varphi$, only $T_1$
executes a write suffix.
Its stores have a fixed program order, so this assignment
contributes exactly one Mazurkiewicz class.

If an assignment satisfies $\varphi$, both threads execute
$m$ stores to $z$.
Their program orders permit exactly
$K=\binom{2m}{m}$ merges of the two store sequences.
Different merges order some pair of conflicting stores
differently and therefore belong to different
Mazurkiewicz classes.
Every merge is realizable by first completing the assignment
pairs as above and then scheduling the two suffixes according
to that merge.
There are no further cross-thread dependencies, so this
assignment contributes exactly $K$ Mazurkiewicz classes.

It follows that
\[
  C_{\mathrm{Maz}}(P_\varphi)
  =(N-s)+Ks
  =N+(K-1)s,
\]
and hence
\begin{equation}
\label{eq:app-ratio-sat-counts}
  \rho(P_\varphi)=\frac{N}{N+(K-1)s}.
\end{equation}

\emph{Separating satisfiable and unsatisfiable formulas.}
The amplification satisfies
\[
  K=\binom{2m}{m}
   =\prod_{i=1}^{m}\frac{m+i}{i}
   \ge 2^m
   =16N.
\]
If $\varphi$ is unsatisfiable, then $s=0$ and
$\rho(P_\varphi)=1$.
If $\varphi$ is satisfiable, then $s\ge1$, so
\[
  \rho(P_\varphi)
  \le \frac{N}{N+K-1}
  \le \frac1{16}.
\]
Thus, the polynomial-time construction gives the gap
\[
  \varphi\notin\mathrm{SAT}
    \implies \rho(P_\varphi)=1,
  \qquad
  \varphi\in\mathrm{SAT}
    \implies \rho(P_\varphi)\le\frac1{16}.
\]

Suppose that an FPRAS for $\rho$ exists.
Run it on $P_\varphi$ with relative-error parameter
$\varepsilon=1/4$.
With probability at least $3/4$, its output $\widehat{\rho}$
satisfies
\[
  (1-\varepsilon)\rho(P_\varphi)
  \le \widehat{\rho}
  \le (1+\varepsilon)\rho(P_\varphi).
\]
On this event, an unsatisfiable formula yields
$\widehat{\rho}\ge3/4$, whereas a satisfiable formula yields
$\widehat{\rho}\le5/64$.
Comparing the output with $1/2$ therefore gives a
polynomial-time randomized SAT decision procedure
with bounded error.

To obtain the one-sided error required for $\mathrm{RP}$,
apply the standard SAT witness self-reduction.
For an input formula with $n\ge1$ variables, amplify the
decision procedure so that each call has error probability
at most $1/(4n)$.
Maintain a partial assignment and, for each next variable,
test whether setting it to zero leaves a satisfiable formula.
Choose zero if the test answers yes and one otherwise.
Use fresh randomness for each call, and finally verify
the resulting complete assignment against the original formula.
Constant formulas can be evaluated directly.

If the original formula is unsatisfiable, verification always
rejects.
If it is satisfiable and every decision call is correct,
the procedure preserves satisfiability at every step and
returns a satisfying assignment.
There are at most $n$ calls; their conditional error bounds
and a union bound give success probability at least $3/4$.
Amplification and all reduced program constructions take
polynomial time.
Thus $\mathrm{SAT}\in\mathrm{RP}$, which implies
$\mathrm{NP}=\mathrm{RP}$.
\end{proof}

\section{Finite-Sample Accuracy}
\label{app:estimation-variance}

Strong consistency guarantees convergence as the number
of trials grows, but does not specify the accuracy after
a given number of trials.
We study this question by bounding the probability of
a large relative error after $N$ trials.

Fix a program $P$ and write
\[
  M=C_{\Maz}(P)>0,
  \qquad
  R=C_{\RF}(P)>0,
  \qquad
  \rho=\frac{R}{M}.
\]
Let $(X_i,Y_i)$ be independent repetitions of the paired
estimator.
The contributions are nonnegative and have finite
second moments, with
\[
  \mathbb E[X_i]=M,
  \qquad
  \mathbb E[Y_i]=R.
\]
Write
\[
  \overline X_N=\frac1N\sum_{i=1}^{N}X_i,
  \qquad
  \overline Y_N=\frac1N\sum_{i=1}^{N}Y_i.
\]
The ratio estimator is
$\widehat\rho_N=\overline Y_N/\overline X_N$
when $\overline X_N>0$, and zero otherwise.

Our analysis considers the two count estimates together.
If both estimates are $20\%$ too large, for example,
their ratio is still correct.
More generally, their errors cancel when they
overestimate or underestimate the counts by similar
proportions.
We measure the error that remains through
$Y_i-\rho X_i$.

\begin{proposition}[Finite-sample accuracy]
\label{prop:ratio-finite-sample}
Write
\[
  V=\operatorname{Var}(Y_i-\rho X_i).
\]
For every $N\geq1$ and $0<\varepsilon<1$,
\[
  \Pr\!\left(
    |\widehat\rho_N-\rho|\geq\varepsilon\rho
  \right)
  \leq
  \min\!\left\{
    1,\,
    \frac{4\operatorname{Var}(X_i)}{NM^2}
    +
    \frac{4V}{N\varepsilon^2R^2}
  \right\}.
\]
\end{proposition}

The first term bounds the probability that the estimated
Mazurkiewicz count is less than half its true value.
Dividing by a small estimate can amplify the error
in the ratio.
The second term accounts for the cancellation between
the two count estimates.
Indeed,
\[
  \frac{Y_i-\rho X_i}{R}
  =
  \frac{Y_i-R}{R}
  -
  \frac{X_i-M}{M}.
\]
This is exactly the difference between the two relative
errors.
Its expectation is zero, and $V/R^2$ measures how much
it varies.
When the relative errors tend to stay close, this
variance is small, and fewer trials suffice to control
this part of the error.

The bound gives a sufficient sample budget for a desired
relative accuracy and failure probability when the true
counts and variances are known.
These quantities are generally unknown; substituting
sample estimates for them does not by itself preserve
the guarantee.

\begin{proof}
Write
\[
  D_i=Y_i-\rho X_i,
  \qquad
  \overline D_N=\frac1N\sum_{i=1}^{N}D_i.
\]
Since $\rho=R/M$, we have $\mathbb E[D_i]=0$.
Independence between trials gives
\[
  \operatorname{Var}(\overline X_N)
  =\frac{\operatorname{Var}(X_i)}{N},
  \qquad
  \operatorname{Var}(\overline D_N)=\frac{V}{N}.
\]

On the event $\overline X_N\geq M/2$, the denominator
is positive and
\[
  |\widehat\rho_N-\rho|
  =
  \frac{|\overline Y_N-\rho\overline X_N|}
       {\overline X_N}
  =
  \frac{|\overline D_N|}{\overline X_N}
  \leq
  \frac{2|\overline D_N|}{M}.
\]
Thus, a relative error of at least $\varepsilon$
requires either $\overline X_N<M/2$ or
$|\overline D_N|\geq\varepsilon R/2$.
Consequently,
\[
  \begin{aligned}
    \Pr\!\left(
      |\widehat\rho_N-\rho|\geq\varepsilon\rho
    \right)
    &\leq
      \Pr(\overline X_N<M/2)
      +
      \Pr\!\left(
        |\overline D_N|\geq\varepsilon R/2
      \right)\\
    &\leq
      \frac{4\operatorname{Var}(X_i)}{NM^2}
      +
      \frac{4V}{N\varepsilon^2R^2},
  \end{aligned}
\]
where the second inequality follows from Chebyshev's
inequality.
Zero-denominator outcomes are included in the first event.
Taking the minimum with one proves the proposition.
\end{proof}

\end{document}